\documentclass[journal=nalefd,manuscript=letter]{achemso}

\setkeys{acs}{usetitle=true}
\setkeys{acs}{doi=true}

\usepackage[version=3]{mhchem} % Formula subscripts using \ce{}
\usepackage{comment}
\usepackage{booktabs}
\usepackage{makecell}

\usepackage{color} 
\usepackage{xr}
\makeatletter
\newcommand*{\addFileDependency}[1]{% argument=file name and extension
	\typeout{(#1)}
	\@addtofilelist{#1}
	\IfFileExists{#1}{}{\typeout{No file #1.}}
}
\makeatother

\newcommand*{\myexternaldocument}[1]{%
	\externaldocument{#1}%
	\addFileDependency{#1.aux}%
}
\myexternaldocument{SI}
\author{Adithya Sadanandan}
\email{adithyasadanandan@ku.edu}
\affiliation[University of Kansas]
{Department of Physics and Astronomy, University of Kansas, Lawrence, Kansas 66045, United States}
\author{Tyson Karl}
\affiliation[University of Kansas]
{Department of Physics and Astronomy, University of Kansas, Lawrence, Kansas 66045, United States}
\author{Rahil Shaik}
\affiliation[University of University of Florida]
{Department of Physics, University of Florida, Gainesville, Florida 32611, United States}
\author{Qunfei Zhou}
\affiliation[University of Kansas]
{Department of Physics and Astronomy, University of Kansas, Lawrence, Kansas 66045, United States}

\email{qunfei.zhou@ku.edu}
\title[An \textsf{achemso} demo]
  {Collective Electrostatics and Band Alignment in Janus MoSTe nanotubes}
\begin{document}
\externaldocument{output}
%%%%%%%%%%%%%%%%%%%%%%%%%%%%%%%%%%%%%%%%%%%%%%%%%%%%%%%%%%%%%%%%%%%%%
%% The "tocentry" environment can be used to create an entry for the
%% graphical Table S1of contents. It is given here as some journals
%% require that it is printed as part of the abstract page. It will
%% be automatically moved as appropriate.
%%%%%%%%%%%%%%%%%%%%%%%%%%%%%%%%%%%%%%%%%%%%%%%%%%%%%%%%%%%%%%%%%%%%%
\begin{tocentry}

%%Some journals require a graphical entry for the Table S1of Contents.
%%%%text is correct.

%Inside the \texttt{tocentry} environment, the font used is Helvetica
%8\,pt, as required by \emph{Journal of the American Chemical
%Society}.

%The surrounding frame is 9\,cm by 3.5\,cm, which is the maximum
%permitted for  \emph{Journal of the American Chemical Society}
%graphical Table S1of content entries. The box will not resize if the
%content is too big: instead it will overflow the edge of the box.

%This box and the associated title will always be printed on a
%separate page at the end of the document.

\includegraphics[width=9cm,height=3.5cm,keepaspectratio]{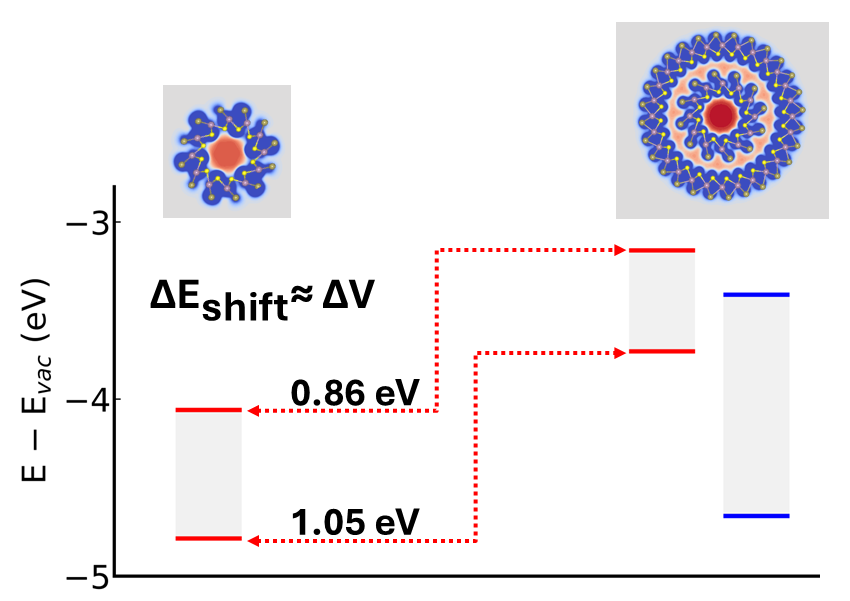}

\end{tocentry}

%%%%%%%%%%%%%%%%%%%%%%%%%%%%%%%%%%%%%%%%%%%%%%%%%%%%%%%%%%%%%%%%%%%%%
%% The abstract environment will automatically gobble the contents
%% ifthe target journal does not use an abstractl.
%%%%%%%%%%%%%%%%%%%%%%%%%%%%%%%%%%%%%%%%%%%%%%%%%%%%%%%%%%%%%%%%%%%%%
\begin{abstract}
	
In this work, we investigate the collective electrostatic effects of one-dimensional (1D) Janus MoSTe nanotubes and their impacts on the band alignment of nanotube heterostructures. Using first-principles calculations based on Density Functional Theory, we find that the Janus nanotube generates a large and uniform electrostatic potential of over 1.3 V within the nanotube pores, which is accumulative for double wall nanotubes. We develop an analytical model to provide a quantitative understanding of the electrostatic potential and its dependence on the quadrupole moment and nanotube radius. For double wall MoSTe nanotube, we find a substantial band edge shift of about 1.0 eV for the inner tube originated from the electrostatic effects, leading to a type-II band alignment. These results demonstrate that the electrostatic effects of 1D nanotubes can be used to tune the electronic properties and band alignment of 1D nanotube heterostructures for optoelectronic and catalytic applications.
 
\end{abstract}
%%%%%%%%%%%%%%%%%%%%%%%%%%%%%%%%%%%%%%%%%%%%%%%%%%%%%%%%%%%%%%%%%%%%%
%% Start the main part of the manuscript here.
%%%%%%%%%%%%%%%%%%%%%%%%%%%%%%%%%%%%%%%%%%%%%%%%%%%%%%%%%%%%%%%%%%%%%
%\section{Introduction}

\newpage

Two-dimensional (2D) transition metal dichalcogenides (TMDs), such as MoS$_2$, WSe$_2$, and MoTe$_2$, etc., have demonstrated intriguing electronic and optical properties promising for electronic and optoelectronic applications, and advancing further device miniaturization \cite{chhowalla2013chemistry, fang2020recent, Lei2022_Graphene}. Janus TMDs, an emerging class of TMDs, have been successfully synthesized by substituting a different chalcogen atom on one side of the TMD\cite{lu2017janus}, leading to an out-of-plane mirror symmetry breaking, therefore an intrinsic out-of-plane polarization\cite{cheng2023effect}.  %In contrast to traditional TMDs, Janus TMDs have a built-in electric field resulting from the intrinsic out-of-plane dipole moment \cite{cheng2023effect}. 
The asymmetrical structure makes Janus structures a compelling platform to explore Rashba spin-splitting and piezoelectric effects \cite{yao2017manipulation,hu2018intrinsic,dong2017large,ahammed2020ultrahigh}. While 2D Janus TMDs show extraordinary %polarity-driven
properties, wrapping Janus TMDs into nanotubes introduces additional symmetry reduction and curvature, further enriching the electrostatic and electronic properties\cite{chowdhury2020progress,zhao2023curvature}. Experimental synthesis of the Janus MoSSe nanotubes has been reported\cite{yang2025janus,lu2017janus}. Theoretical studies on Janus TMDs suggest that MoSTe nanotubes with a few-nanometer radius show promising thermal stability comparable to their parent 2D structures\cite{bolle2021structural}. Moreover, phonon calculations predict the dynamic stability of MoSTe monolayers and nanotubes, making it a suitable representative candidate for exploring electronic properties of Janus TMD nanotubes \cite{yagmurcukardes2019electronic, zhao2015ultra}. %Recently, Yang \textit{et al.} \cite{yang2025janus} reported the first experimental realization of Janus MoSSe nanotubes , boosting the study of nanotubes for further exploration. \par 

When Janus TMD nanosheets are wrapped into 1D nanotubes, the charge distribution is significantly altered, arranging the out-of-plane dipoles from 2D periodic arrays to a radial distribution arranged periodically along the tube axis. As highlighted by Zojer\cite{zojer2024electrostatically}, %,zojer2023electrostatic,zojer2024tuning,zojer2019impact,zojer2025collective}, 
the electrostatic response of the dipoles arranged in different periodic configurations gives rise to various intriguing collective electrostatic effects. The electric field from these electrostatic effects can remarkably modify the energy landscape of low-dimensional materials, tuning their electronic and optical properties \cite{kraberger2015tuning,Chan2024NL}. Organic molecular layers with out-of-plane dipoles can significantly change the work function and band edge energies for bulk surfaces\cite{Farber2025BITC} and 2D materials\cite{Zhou2019NL,zojer2019impact,zojer2018DipolarAdvFuncMater,zojer2018Dipolar2DMater}. Self-assembled organic molecular layers with negligible out-of-plane dipole moment but quadrupole moments can generate periodic electrostatic potentials that modulates the electronic\cite{zhou2021engineering} and excitonic\cite{Chan2024NL} properties of 2D materials and their heterostructures. The collective electrostatics in porous metal-organic and covalent organic frameworks\cite{zojer2023electrostatic,zojer2025collective,zojer2024tuning,zojer2020MOF,zojer2021MOFAdvMater} with cylindrical arrangements of dipolar units give rise to intriguing electrostatic energy within the pores that are highly tunable through functionalization of their building blocks.
%\textcolor{green}{ref: https://doi.org/10.1002/admi.201500323}.%, 
%While the electronic structures of Janus TMD nanotubes have been reported, %with reduced band gaps for smaller nanotubes. This is contradictory to quantum confinement effects, due to dominant curvature-induced flexoelectricity and strain \cite{zhao2023curvature, bennett2021flexoelectric,mikkelsen2021band}. %resulting from curvature-induced flexoelectricity and strain \cite{zhao2023curvature, bennett2021flexoelectric,mikkelsen2021band}, dominant over quantum confinement effects. 
%these studies focus mainly on the effects of strain and curvature on the band gap and band alignment\cite{zhao2015ultra, mikkelsen2021band,luo2019electronic, lan2023structural}. 
Studies on the electronic properties of Janus TMD nanotubes have been mainly focused on effects of strain and curvature\cite{zhao2015ultra, mikkelsen2021band,luo2019electronic, lan2023structural}. 
However, the electrostatic effects in the Janus TMD nanotubes are largely underexplored. A combined classical electrostatics model with Bader analysis predicts that the diameter-dependent electric field regulates the band alignment for TMD heterotubes \cite{ge2024band}. Understanding the electrostatic effects of Janus TMD nanotubes and their influence on the band structure and interface properties is essential for the design of 1D nanotubes for applications in photovoltaics\cite{cai2023ultrafast}
%\textcolor{green}{add ref: https://doi.org/10.1103/PhysRevB.108.045416},
%photocatalytic water splitting, 
catalysis, piezoelectric, and spintronic devices \cite{yin2021recent, zhao2025preparation}.  

In this work, we investigate the collective electrostatic effects for 1D MoSTe nanotubes of various sizes using first-principles calculations based on Density Functional Theory (DFT). Using a discretized charge density (DCD) model and derivation of an analytical formula, we unravel the direct relationship between the diameter-dependent electrostatic potential from the nanotube and the quadrupole moment arising from the radially distributed dipoles. This model works for both single wall (SW) and double wall (DW) nanotubes. Local density of states shows a large band edge shift of about 1.0 eV for the inner tube of a DW nanotube, leading to a type-II band alignment for the 1D van der Waals heterostructure. This large band edge shift is equivalent to the interface-modulated electrostatic potential at the center of the outer nanotube, indicating an electrostatic physical origin. These findings demonstrate that electrostatic potential in 1D nanotubes offers a versatile strategy to tune the electronic states and band alignment for 1D nanotubes and their heterostructures.
% works for other 1D Janus TMD nanotubes as well.

\begin{figure}
    \centering
    \includegraphics[width=0.5\linewidth]{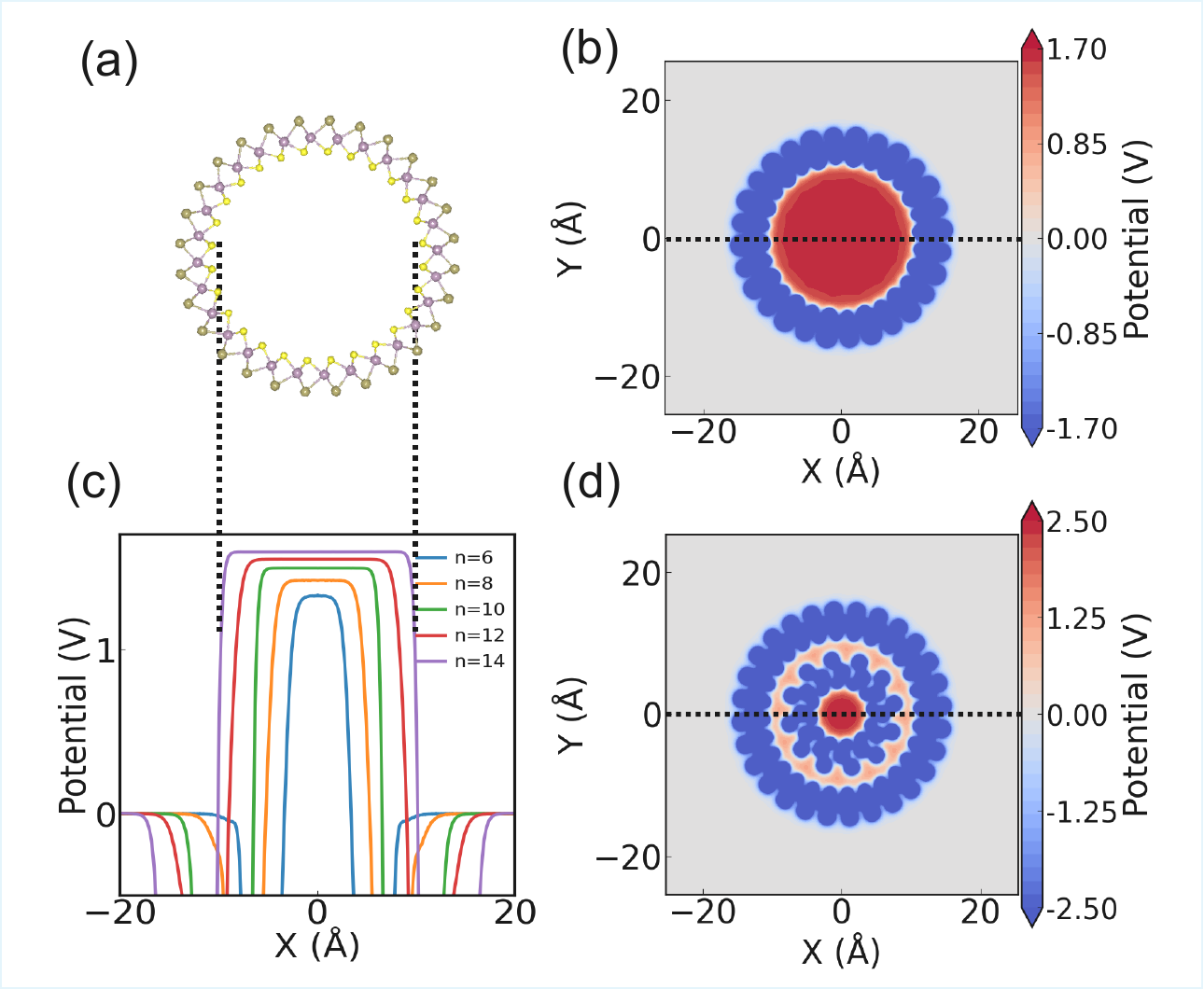}
    \caption{The atomic structure (a) and real-space electrostatic potential at the $z=1.0$ \AA{} plane for (b) SW MoSTe nanotube ($n = 14$), and (d) DW MoSTe nanotube. (c) Electrostatic potential of the SW MoSTe nanotubes along the $y = 0$ \AA{} line as labeled in (b). All potentials are referenced to the vacuum potential. $z$-axis is along the tube axis. The electrostatic potential in this work corresponds to the potential experienced by an electron.}
%    \textcolor{red}{can you plot with the same $z$ for both SW and DW?}\textcolor{blue}{both plots uses Z= 1 \AA{} now}}
    \label{fig:1}
\end{figure}

%\section{Results and discussion}
The initial structures for MoSTe nanotubes are adopted from the DTU data repository with sizes of ($n$= 6,8,10,12,14) in the armchair direction \cite{bolle2021structural, lan2023structural}, where $n$ is the number of unit cell repetitions. The MoSTe nanotube has a more stable structure with S atoms at the inner side of the tube and Te atoms outside, as shown in Fig.~\ref{fig:1}(a). The structures have been further optimized with DFT in this work. The atomic structures and computational details are provided in the supplementary information (SI). 
 
%Here, the long axis of the nanotube is aligned with the $z$ axis. 
The electrostatic potentials, defined as the potential of an electron in this work, for the MoSTe nanotubes are shown in Fig.~\ref{fig:1} and Fig. S4. The tube axis is aligned with the $z$ axis. As shown in Fig.~\ref{fig:1}(b-c), the MoSTe nanotube generates a uniform electrostatic potential inside the nanotube, which changes negligibly with $z$, see Fig. S4(e-h) and Fig. S5(b). The center electrostatic potential is positive compared with the vacuum level outside of the nanotube, and increases with the increase of nanotube radius, from 1.31 V for $n=6$ to 1.60 V for $n=14$ MoSTe nanotube. For the DW nanotube that consists of $n=6$ and $n=14$ nanotubes, the electrostatic potential inside the inner nanotube is contributed from the two nanotubes, with an increased value of about 2.41 V.%, though slightly smaller than the sum of the two nanotubes.%, as a result of interfacial charge redistribution, as shown in Fig.~\ref{fig:S3}. 

\begin{figure}
    \centering
    \includegraphics[width=0.5\linewidth]{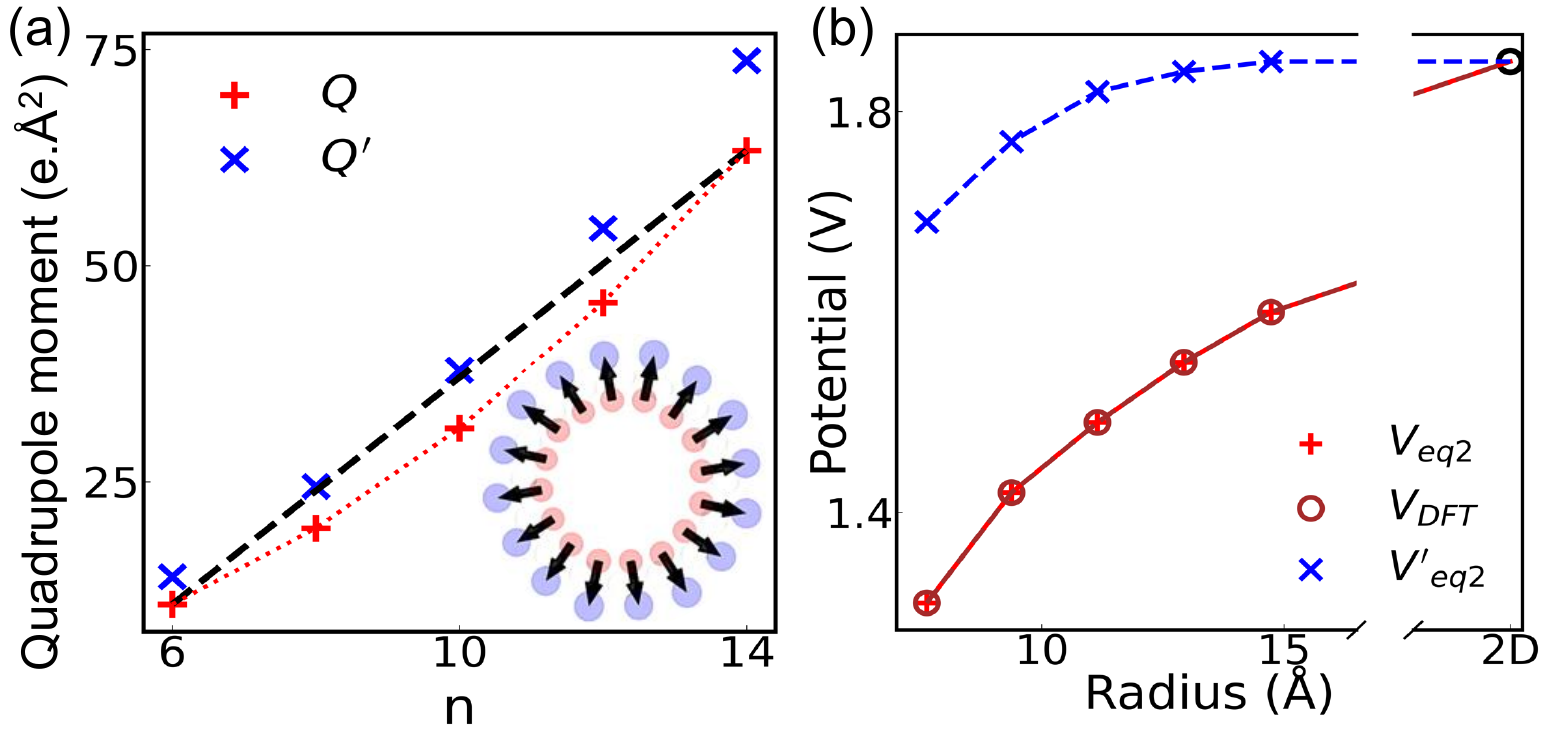}
    \caption{The quadrupole moment (a), and the electrostatic potential from DFT ($V_{DFT}$) and from Eq.~\ref{eq:V2} ($V_{eq2}$) (b). Here, $Q$ and $V_{eq2}$ are computed using the radius-dependent effective charges as shown in Table S1. For comparison, the quadrupole moment and electrostatic potential for all nanotubes are also calculated using a common charge $q = 0.0298 e$, which are denoted as $Q'$ in (a) and $V'_{eq2}$ in (b), respectively. $q = 0.0298 e$ is the effective charge for 2D MoSTe monolayer. 
    %Here, $Q$ and $V_{eq2}$ are computed using the radius-dependent effective charges as shown in Table S1.} For comparison, the quadrupole moment and electrostatic potential are also calculated using \textcolor{blue}{a common charge $q = 0.0298 e$}, which is the effective charge for 2D MoSTe monolayer, and are denoted as $Q'$ in (a) and $V'_{eq2}$ in (b), respectively. 
    The inset in (a) is the nanotube structure with Mo (white), Te (blue), and S (red) atoms, with dipoles pointing from S to Te. The black dashed line connects the first and last $Q$ values. All the other dashed lines are just connecting the data points used as aid of the eye. For comparison, the electrostatic potential difference between the two sides of the 2D MoSTe monolayer (Fig. S7) is included in (b), assuming it’s a tube with infinite radius.}        
%    	(a) The quadrupole moment $Q$ from $q$ for each size $n$ (red) and $Q'$ calculated using the same $q=0.0298$ of 2D MoSTe (blue), for SW MoSTe. % as a function of nanotube radius.
%    The red dotted line connects the individual $Q$ values, whereas the black dashed line is the reference line connecting the first and last $Q$ values, highlighting the overall increasing trend. The inset in (a) is the nanotube structure with atoms Mo (white), Te (blue), and S (red), and the dipoles pointing from S to Te atoms.
%    (b) The electrostatic potential from DFT ($V_{DFT}$), from Eq.~\ref{eq:V1} with the corresponding $q$ for each radius($V_{eq1}$) and that with the same $q$ value of 2D MoSTe ($V'_{eq1}$). The lines connecting the data points are included to illustrate the convergence towards the 2D limit. The 2D limit value corresponds to the DFT calculated potential for the MoSTe monolayer.   
    \label{fig:2}
\end{figure}

To obtain a better understanding of this radius-dependent electrostatic potential, we model the electrostatics of the 1D nanotube as attributed from radially distributed dipoles, pointing from the S to Te atoms and arranged periodically along the nanotube axis, see inset in Fig.~\ref{fig:2}(a). We use a discretized charge density (DCD) model\cite{zhou2021engineering,zhou2022analytical} to represent these radially distributed dipoles using effective positive and negative point charges located at the positions of the Te and S atoms, respectively. In this work, we follow the physics convention and define the direction of dipoles pointing from negative to positive charges. This DCD model is similar to the Density Derived Electrostatic and Chemical charges (DDEC) \cite{manz2012improved} with the purpose of reproducing the electrostatic potential outside of the electron distribution of the material. The difference is that the point charges in the DCD model are not the net atomic charge for each atom as in the DDEC model, but only to obtain an effective description of the dipoles and quadrupoles for the material. As in 1D Janus nanotubes, only two point charges are used to represent each of the radially distributed dipoles, without considering the Mo atoms.
%e.g., in the case of 1D Janus nanotubes, the Mo atoms are not considered. %using only two point charges to represent each of the radially distributed dipoles. 
There are also many other methods for determining atomic charges, such as the Hirshfeld charges \cite{hirshfeld1977bonded}, Charge Model 5 (CM5) \cite{marenich2012charge}, Bader charge\cite{henkelman2006fast}, and Mulliken Population analysis \cite{mulliken1955electronic}, which provide different atomic charge values, used for different material/molecular properties analysis. The DCD model representation for the 1D nanotube is different from these methods, and is only used to describe and better understand the electrostatic potential.

With the DCD model, we can calculate the electrostatic potential for the nanotube.
Mellin transform and Poisson resummation are used to account for the 1D periodic charge distribution, ensuring faster convergence in the reciprocal space\cite{zhou2021engineering,zhou2022analytical}. The potential at position (R, z) outside of the electron density of the nanotube is given by (See SI for details of the derivation )

\begin{equation}
V(R,z) = \frac{2K}{c}
\sum_j q_j \left(
-\ln R_j
+ \sum_{k \neq 0} \frac{1}{\pi}
\exp\!\left(
-2\pi i k \frac{z - z_j}{c}
\right)
\kappa_0\!\left(
2\pi |k| \frac{R_j}{c}
\right)
\right)
\label{eq:V1}
\end{equation}

Here $K, c, \kappa_0$ are the Coulomb's constant, lattice parameter, and the modified Bessel function of the second kind, respectively. $R$ is the radial distance of the subject point to the center axis of the tube. $R_j$ is the radial distance of charge $q_j$ from the subject point. $\sum_j q_j = 0$ ensures charge neutrality.%The radial distance $R_j$ is given by $\sqrt{(x-xj)^2+(y-y_j)^2}$. 

%The dimensionless axial coordinate, $\delta= (z-z_j)/c$.
%where $(x,y,z)$ are the coordinates of the evaluation point, the center of the nanotube, and $(x_j, y_j, z_j)$ are the coordinates of the S and Te atoms in the DFT-relaxed structure}\textcolor{green}{does Eq.1 work for any (x,y,z) or just the center? %$R_j=\sqrt{(xj)^2+(y_j)^2}$ is the vertical distance to the tube center axis ($x=y=0$) for charge $q_j$.
%In the DCD model, positive and negative charges reside at the positions of the Te and S atoms, respectively, with the same absolute charge values. 

With Eq.~\ref{eq:V1}, we obtain the effective charge $q$ values in the DCD model by fitting the electrostatic potential with DFT-calculated results, see details in the SI. The $q$ value increases with larger nanotube size $n$, see Table S1, approaching the value of 0.0298$e$ for 2D MoSTe, which is obtained by fitting the polarization from the DCD model with that from DFT, see details in the SI. %This indicates that when rolled into 1D, local dipole-dipole interactions induce a depolarization for the radial dipoles, which is enhanced for smaller radii. %curvature-induced charge redistribution
This indicates that the radial polarization is curvature-dependent.

While the net total charge and dipole moment is zero, the electrostatic potential is a result of the quadrupole moment of the 1D charge distribution. The quadrupole moment $Q$ can be simplified as \(Q=\sum_j q_jR_j^2=qN(R_{Te}^2 - R_S^2)\), where $N, R_{Te}, R_S$ are the number of dipoles, the radial distance of the Te and S atoms from the nanotube center, respectively. With this, Eq.~\ref{eq:V1} can be further derived as  (details are included in SI),

\begin{equation}
    V = \frac{2KQ}{cd(D-d)}\ln{\frac{D}{D-2d}} = \frac{2KpN}{cd}\ln{\frac{D}{D-2d}}
    \label{eq:V2}
\end{equation}
where $D = 2R_{Te}$ is the outer diameter of the tube, and $d = (R_{Te}- R{s})$, the radial thickness of the nanotube, $p=qd$ the dipole moment of the single $S-Te$ pair in the nanotube. % \textcolor{blue}{the Eq. \ref{eq:V2} can be further expressed as,
%\begin{equation}
%    V  = \frac{2KpN}{cd}\ln{\frac{D}{D-2d}}
%    \label{eq:V3}
%\end{equation}} 
%For DW nanotubes, one effective charge value is used for both the inner and outer tubes, which have different diameter $D$. Therefore the electrostatic potential for DW tubes can be calculated with Eq.~\ref{eq:V2} by summing up the contributions from the inner and outer tubes. 
%For DW, Eq.~\ref{eq:V2} is evaluated separately for inner and outer tubes, and the total potential is obtained by adding their contributions. 
%\textcolor{green}{for the DW, can you use D1 and D2, the outer diameter of inner and outer tube, instead of just one D? then you will have two first-order term? if D is infinite, does V approach to that for 2D?}
%With only the first-order approximation, the electrostatic potential from Eq.~\ref{eq:V1} and Eq.~\ref{eq:V2} agrees very well, see Table S1. 
As shown in Eq.~\ref{eq:V2}, the electrostatic potential is strongly dependent on the quadrupole moment $Q$, or the dipole moment $p$ of the radial dipoles, and nanotube size. 
%\textcolor{green}{move column $V_{eq2}$ before column $V'_{eq1}$ in Table S1.}

%\vspace{-10pt}

The electrostatic potential from Eq.~\ref{eq:V2}, %(\textcolor{red}{or Eq.~(\ref{eq:V1})? I put eq1 for $V_{eq1}$ as there's $q$ in it. please change it if not)}(\textcolor{blue}{(Electrostatic potential is from eq1)}, 
denoted as $V_{eq2}$, agrees well with that from DFT, $V_{DFT}$, as shown in Fig.~\ref{fig:2} and Table S1. This indicates the validity of Eq.~\ref{eq:V1}–\ref{eq:V2}, with which we can further investigate the impacts of the curvature-induced depolarization %\textcolor{blue}{geometric effects} 
on the quadrupole moment and electrostatic potential. By definition, the quadrupole moment depends linearly on the number of dipoles $N=2n$ for each unit cell of the nanotube. 
However, the quadrupole moment $Q$ curve lies below the linear line connecting the two end $Q$ values, see Fig.~\ref{fig:2}(a), indicating the depolarization effects, same as that shown for the $q$ values, which decreases as the tube radius decreases.
%However, the quadrupole moment $Q$ curve lies below the linear line connecting the two end $Q$ values, see Fig.~\ref{fig:2}(a), indicating \textcolor{blue}{curvature-induced reduction of $Q$}, same as that shown for the $q$ values, which enhances as the tube radius decreases.
%\textcolor{green}{put $(R_{Te}^2 - R_S^2)$ value in Table S1? } included 

%\textcolor{blue}{This reduction of $Q$ and effective charges with larger curvature can be understood by space-modulated accumulation of charges. For MoSTe nanotubes, the S atoms at the inner side of the tube, is more electronegative and accumulate larger electron densities than for Te atoms at the outer side of the tube, resulting in a dipole pointing from S to Te radially. When nanotube radius is reduced, the nearest S-S (Te-Te) distance decreases (increases) from about 3.05 (3.70) \AA\ for $n=14$ to about 2.82 (4.30) \AA\ for $n=6$, less number of electron density can be accumulated around S atoms. This leads to a reduction of the effective charge value and quadrupole moment.}
This quadrupole moment and effective charge reduction with larger curvature can be understood by space-modulated charge accumulation. For MoSTe nanotubes, the S atoms at inner side of the tube, is more electronegative and accumulate larger electron densities than for Te atoms at the outer side, resulting in a dipole pointing from S to Te radially. When nanotube radius is reduced, the nearest S-S (Te-Te) distance decreases (increases) from about 3.05 (3.70) \AA\ for $n=14$ to about 2.82 (4.30) \AA\ for $n=6$. Therefore less electron density can be accumulated around S atoms, leading to a reduction in effective charge and quadrupole moment.
Here we calculate the electrostatic potential (quadrupole moment) in two ways: (1) using the corresponding curvature-dependent effective charge for each nanotube as shown in Table S1, where the interior potential (quadrupole moment) is denoted as $V_{eq2}$ ($Q$); (2) using a common charge $q=0.0298e$ for all nanotubes, denoted as $V'_{eq2}$ ($Q'$). As $q=0.0298e$ is the effective charge for 2D MoSTe monolayer, the difference between $V_{eq2}$ ($Q$) and $V'_{eq2}$ ($Q'$) quantifies the impact of the curvature-dependent depolarization effects. 
%\textcolor{blue}{The electrostatic potential (quadrupole moment) calculated using the corresponding effective charge for each nanotube is denoted as $V_{eq2}$ ($Q$), and that calculated using the common $q=0.0298e$ is denoted as $V'_{eq2}$ ($Q'$).}

%%\textcolor{blue}{This effect}
%The impact of depolarization can be quantitatively captured by using %the same 
%\textcolor{blue}{a common} $q=0.0298e$ value, which is the effective charge for the 2D MoSTe monolayer, to calculate the quadrupole moment and electrostatic potential, denoted as $Q'$ and $V'_{eq2}$, respectively, for all nanotubes. 

% for all nanotubes, and calculating the quadrupole moment and electrostatic potential, denoted as $Q'$ and $V'_{eq1}$, respectively.
%The smaller radius tubes have an enhanced depolarization effect. 
The relative deviation in quadrupole moment, $(Q'-Q)/Q$, and potential difference, $(V'_{eq2}-V_{eq2})$, increase systematically with decreasing nanotube radius, as shown in Table S1, from 16\%  and 0.25 V for $n=14$ to 29\% and 0.38 V for nanotubes $n=6$, respectively.
%\textcolor{blue}{(the change in the quadrupole moment is not unitform, the highest change is for n = 8, lowest for n} 
%\textcolor{green}{$q$ has larger difference for n=6, why n=8 has a larger Q difference?} 
Therefore, the depolarization effect is non-negligible for SW nanotubes of small radius.

 For the DW nanotube, in addition to the intra-nanotube depolarization, charge redistribution at the interface as a result of charge transfer further modulates the quadrupole field. As shown in Fig. S4, significant charge redistribution occurs at the DW nanotube interface, with charge depletion (accumulation) around the Te atoms of the inner tube (S atoms of the outer tube), indicating interfacial electron transfer from the inner to outer tubes. This leads to radially arranged dipoles pointing toward the nanotube center, in opposite directions as those for the inner and outer nanotubes. 
To account for the interface charge transfer, we use one effective charge value for both the inner and outer tubes of the DW nanotube. In this case, the electrostatic potential for DW nanotubes can be calculated with Eq.~\ref{eq:V2} by summing up the contributions from the inner and outer tubes that have different diameters $D$.
 This interface-induced quadrupole moment reduces the total quadrupole moment, leading to a smaller effective charge $q$, and a large difference between $V'_{eq2}$ and $V_{eq2}$ for the DW nanotube, see Table S1. As $V'_{eq2}$ includes no interface effect, its value for the DW nanotube, 3.62 V, is close to the sum of $V'_{eq2}$ for the two isolated $n=6$ and $n=14$ SW nanotubes, 3.54 V. The sum of $V_{eq2}$ for the two component SW nanotubes, 2.91 V, is about 0.50 V larger than $V_{eq2}$ of the DW nanotube, which can be attributed to the interfacial charge transfer . %the interface charge redistribution.
% \textcolor{green}{'Table S1' should be 'Table S1', check all}

%For DW nanotubes, one effective charge value is used for both the inner and outer tubes, which have different diameter $D$. Therefore the electrostatic potential for DW tubes can be calculated with Eq.~\ref{eq:V2} by summing up the contributions from the inner and outer tubes. 
 
 As shown above, the electrostatic potential of the Janus MoSTe nanotube interior is strongly dependent on the quadrupole moment of the nanotube, which can be successfully described as radially arranged dipoles represented by the DCD model and can be largely tuned. The quadrupole moment is dependent on the dipole moment of each dipole, and the depolarization induced by curvature-dependent charge redistribution. %, as well as interface induced charge redistribution. 
 The dipole moment of each dipole can be tuned by the relative electronegativity of the anions, i.e., S and Te atoms for MoSTe, which determines the effective charge $q$ value. While positive electrostatic potentials are obtained for MoSTe, negative potentials can be obtained if the radially-distributed dipoles are pointing toward the center.
% Depolarization is reduced for nanotubes of larger radius, with the dipole moment approaching that of the 2D counterpart. 
The electrostatic potential is accumulative. Therefore larger potentials inside the nanotube can be obtained with DW and multi-wall nanotubes.  %for DW nanotubes, though reduced with interface-induced effects. 
There are over 100 Janus nanotubes theoretically predicted to be stable\cite{bolle2021structural}, and Janus MoSSe nanotubes have been realized experimentally \cite{yang2025janus}. %with radius small enough to exhibit distinctive properties than their corresponding 2D materials. 
%Although only MoSTe nanotubes and one DW nanotube are studied in this work, the same effects and the analytical model from this work applies for other types of nanotubes.
%Meanwhile, though armchair MoSTe nanotubes are mainly studied in this work, 
Same electrostatic effects and Eq. 2 also apply for zigzag MoSTe nanotubes, see details in Fig. S10 and Table S1. This demonstrates that the understanding of electrostatic effects and analytical model derived in this work can be applied for other nanotubes of different compositions and chiralities.
%is universal for 1D nanotubes of different chiralities, and 

\begin{figure}
    \centering
    \includegraphics[width=0.5\linewidth]{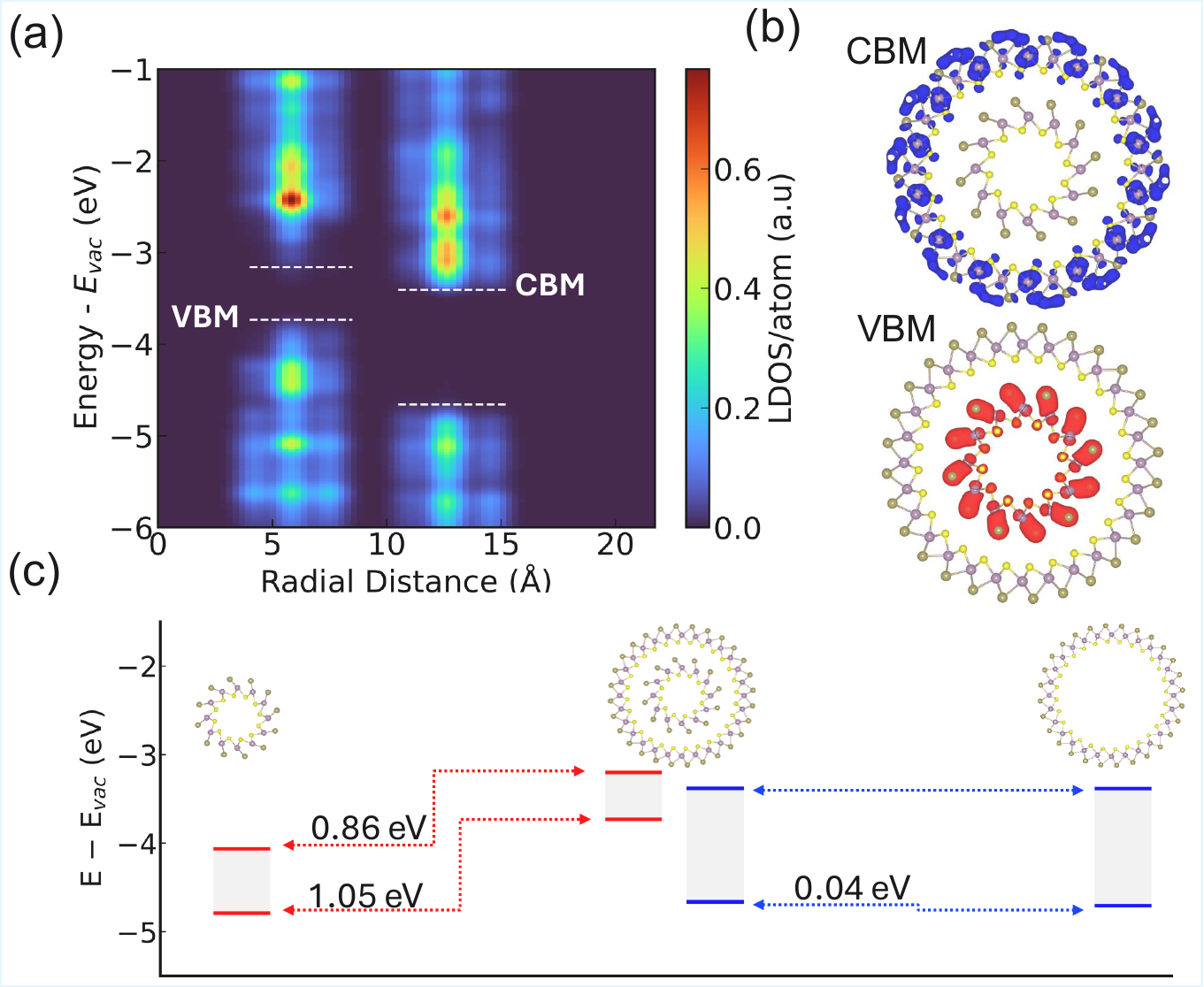}
    \caption{(a) Radially-resolved local density of states (LDOS) of the DW MoSTe, where zero is the center of the nanotube. The dashed horizontal lines show the conduction band minimum (CBM) and valence band maximum (VBM). (b) The real-space CBM and VBM wavefunctions (charge density isosurfaces) of the double wall MoSTe view along the tube axis. The isosurface shown in (b) is the squared modulus of the wavefunction. The isosurface value of the VBM is $2.2\times 10^{-6}$ $e/$\AA$^3$, and that of the CBM is $1.6\times 10^{-6}$ $e/$\AA$^3$. (c) VBM and CBM band energies for the $n=6$ (red) and $n=14$ (blue) SW nanotubes and that in the DW nanotube. All energies are referenced to the vacuum energy.}
    \label{fig:3}
\end{figure}

Finally, we show that the electrostatic potential can be used to tune the band alignment for DW nanotubes. For SW Janus MoSTe nanotubes, the band gap generally decreases for smaller nanotube radius due to curvature-induced band gap reduction, dominant over quantum confinement effects\cite{mikkelsen2021band}. The valence band maximum (VBM) and conduction band minimum (CBM) for $n=6$ and $n=14$ SW MoSTe nanotubes are shown in Fig.~\ref{fig:3}(c), with a band gap of 0.72 eV and 1.32 eV, respectively.% agree well with \cite{mikkelsen2021band}
%For the DW nanotube, local density of states (LDOS) is averaged along to analyze the energy alignment. 

For the DW nanotube, which consists of $n=6$ and $n=14$ nanotubes, the local density of states (LDOS) is calculated and averaged along the radial direction from the center nanotube axis, as shown in Fig.~\ref{fig:3}(a). The LDOS shows clearly the states for the inner tube of radial distance within $\sim$10 \AA\ and the outer tube separated by the interface.
%%%%%%%%%%

%The LDOS in Fig.~\ref{fig:3} shows a type-II band alignment, with the VBM and CBM located at the inner and outer tubes, respectively, which is explicitly demonstrated by the CBM and VBM wavefunctions, Fig.~\ref{fig:3}(b). This type-II band alignment is strikingly different from the level alignment for the intrinsic SW nanotubes, as shown in Fig.~\ref{fig:3}(c). The inner tube undergoes a substantial band energy shift of about 0.90(1.01) eV for the CBM (VBM). As the outer tube generates a positive, uniform electrostatic potential of 1.60 V within its interior, as shown in Fig.~\ref{fig:1} and Table S1. The band energy levels for the inner tube, residing within this electrostatic potential, can therefore be shifted correspondingly. The electrostatic potential of the outer tube, subtracted by the potential reduction due to interface charge redistribution, leads to a net potential of about 0.99 V, as shown in Fig.~\ref{fig:S2}(a) for the potential within the interface of the DW nanotube. This potential from the outer tube is similar to the band edge energy shift of the inner tube.

The LDOS in Fig.~\ref{fig:3} shows a type-II band alignment, with the VBM and CBM located at the inner and outer tubes, respectively, which is explicitly demonstrated by the CBM and VBM wavefunctions, Fig.~\ref{fig:3}(b). %This type-II band alignment is strikingly different from the level alignment for the intrinsic SW nanotubes, as shown in Fig.~\ref{fig:3}(c). 
This type-II band alignment with the VBM of inner tube within the gap of the outer tube is strikingly different from the level alignment for the isolated Janus SW nanotubes, where the CBM of the SW $n=6$ is at a energy within the gap of the $n=14$ SW nanotube, as shown in Fig.~\ref{fig:3}(c).
The inner tube undergoes a substantial band energy shift of about 0.86(1.05) eV for the CBM (VBM). 

%The LDOS in Fig.~\ref{fig:3} shows a type-II band alignment, with the VBM and CBM located at the inner and outer tubes, respectively, which is explicitly demonstrated by the CBM and VBM wavefunctions, Fig.~\ref{fig:3}(b). This type-II band alignment is strikingly different from the level alignment for the intrinsic SW nanotubes, as shown in Fig.~\ref{fig:3}(c). The inner tube undergoes a substantial band energy shift of about 0.90(1.01) eV for the CBM (VBM). 
%As the outer tube generates a positive, uniform electrostatic potential of 1.60 V within its interior, as shown in Fig.~\ref{fig:1} and Table S1. The band energy levels for the inner tube, residing within this electrostatic potential, can therefore be shifted correspondingly. The electrostatic potential of the outer tube, subtracted by the potential reduction due to interface charge redistribution, leads to a net potential of about 0.99 V, as shown in Fig.~\ref{fig:S2}(a) for the potential within the interface of the DW nanotube. This potential from the outer tube is similar to the band edge energy shift of the inner tube.

As the outer tube generates a positive, uniform electrostatic potential of 1.60 V within its interior, as shown in Fig.~\ref{fig:1} and Table S1, and the interface charge redistribution induced a $\sim$ 0.5 V reduction of the potential, the net electrostatic potential within the outer tube is about 1.10 V. The band energy levels for the inner tube, residing within this electrostatic potential, can therefore be shifted correspondingly, as shown in details in Fig. S9, and in Fig.~\ref{fig:3}(c) for the VBM shift of 1.05 V. %The electrostatic potential of the outer tube, subtracted by the potential reduction \textcolor{blue}{of 0.5 V} due to interface charge redistribution, leads to a net potential of \textcolor{blue}{about 1.10 V inside the outer wall nanotube, which is close to the VBM shift of the inner tube, see Fig.~\ref{fig:3}(c). The energy shift of CBM for the inner tube is smaller than that for the VBM, due to hybridization }
%as shown in Fig.~\ref{fig:S2}(a) for the potential within the interface of the DW nanotube. This potential from the outer tube is similar to the band edge energy shift of the inner tube.
%The VBM shift of 1.05 V for the inner tube, see Fig.~\ref{fig:3}(c), is similar to the electrostatic potential within the interior of the outer tube. 
The band gap of the inner tube is reduced slightly by about 0.19 eV, as a result of electronic hybridization of the states from the inner tube with the conduction bands of the outer tube, see Fig. S9. Therefore the CBM of the inner tube is shifted at a smaller amount of about 0.86 eV than the VBM.

In contrast, there is no consistent band energy shift for the outer tube comparing to the $n=14$ SW nanotube, see Fig.~\ref{fig:3}(c). Instead, it shows a slight reduction in VBM of about 0.04 eV, see Fig.~\ref{fig:3}(c), which is a result of hybridization with the valence bands of the inner tube. The CBM of the outer tube, which lies within the gap of the inner tube, remains unshifted.

%Instead, it shows only a band gap reduction of about 0.07 eV, see Fig.~\ref{fig:3}(c), with CBM (VBM) energies changing by about -0.03 (0.04) eV. This band gap reduction can be attributed to dielectric screening effect. Low-dimensional materials are highly sensitive to the external dielectric environment, which can lead to significant band energy renormalization\cite{gordeev2024dielectric, neaton2006renormalization, zhou2021range}. Properties of nanomaterials encapsulated inside carbon nanotubes have been found to be significantly tuned through dielectric screening by changing the outer tube–the dielectric environment\cite{gordeev2024dielectric}. The dielectric screening effect of the outer tube is slightly larger than that of the inner tube, therefore a slightly larger, 0.15 eV, band gap reduction for the inner tube. This band gap reduction further modifies slightly the band edge shift resulted from the electrostatic potential for the inner tube, leading to a slightly smaller energy shift for the CBM than for VBM,  Fig.~\ref{fig:3}(c).%For the DW, the dielectric screening from the outer tube reduces the band gap of the inner tube by about 0.15 eV. While for the outer tube, the screening effect from the inner tube is negligible, reducing the band gap by only about 0.07 eV.

While the electrostatic potential from the inner tube outside of itself is zero, no electrostatic effects are observed for the outer tube, which is consistent with the unshifted band energies for the outer tube. This further implies that the band energy shift for the inner tube is purely an electrostatic effect of the outer tube. % consistent for both the occupied and unoccupied bands. 

For comparison, we studied the electrostatic potential and band alignment for intrinsic MoS$_2$ 1D nanotubes, see details in Fig. S11, Fig. S12 in the SI. MoS$_2$ nanotubes also generates uniform electrostatic potential within the pores, but in much smaller magnitude than for Janus MoSTe nanotubes due to the lack of intrinsic radial polarization. Despite of this smaller magnitude, the electrostatic potential within the pores shifts the band edge energies correspondingly in DW MoS$_2$ nanotubes, same as that in Janus nanotubes.
%the VBM (CBM) energy shift for the outer tube is only about -0.03 (0.04) eV, which agrees with the band gap reduction of about 0.07 eV, resulting from the dielectric screening effect of the inner tube. While the electrostatic potential outside of the tube is zero, no electrostatic effects are observed for the outer tube. This further implies that the band energy shift for the inner tube is purely an electrostatic effect from the outer tube, consistent for both the occupied and unoccupied bands. \par

%The band gap reduction of about 0.15 eV further modifies the band energy shifts slightly. This indicates that the significant band energy shift for the inner tube originates from the electrostatic potential of the interior of the outer tube.

%In contrast, the VBM (CBM) energy shift for the outer tube is only about -0.03 (0.04) eV, which agrees with the band gap reduction of about 0.07 eV, resulting from the dielectric screening effect of the inner tube. While the electrostatic potential outside of the tube is zero, no electrostatic effects are observed for the outer tube. This further implies that the band energy shift for the inner tube is purely an electrostatic effect from the outer tube, consistent for both the occupied and unoccupied bands. \par

The uniform electrostatic potentials within the nanotube and their impacts on the band alignment indicate the opportunities of using electrostatic engineering to tune properties of nanomaterials and molecules positioned at the interior of the tube, and band alignment for nanotube heterostructures. 
The type-II band alignment, as shown in the DW Janus MoSTe nanotube, can facilitate efficient charge separation, which is essential for photovoltaic %and photocatalytic 
applications. 
%The theoretical prediction of long-lived excitons in MoSTe monolayers \cite{jin2018prediction} and our observation of spatially separated VBM and CBM for the double-wall indicate that Janus TMD nanotubes can be further studied for light harvesting applications. 
%By changing the direction and magnitude of the radially distributed dipoles determined by the relative electronegativity of the surface atoms of the nanotube, the electrostatic potential inside the nanotube can be modified, and engineered to be utilized as a molecular storage medium.
%By combining the confinement effects and electrostatic interactions, the MoSTe nanotube can be utilized as a molecular storage medium. 
%%%
%Encapsulation of molecules is broadly explored within carbon,  BN, and TMD nanotubes \cite{campo2020optical,peng2024nanotube,chen2003titanium}. Inside these nanotubes, the guest molecules experience a modified electronic and electrostatic environment, which subsequently influences the diffusion and stability of the guest molecules. For Janus TMD nanotubes, the intrinsic electrostatic field can effectively create a potential well within the cavity, which can further selectively attract and trap molecules and charged species. 
The electronic energies of organic molecules placed at the pores of the 1D nanotube can also be tuned by the interior electrostatic potential of the nanotube. This has been demonstrated for porous covalent organic frameworks (COFs) where physisorption of C$_{60}$ undergoes a energy level alignment change equivalent to the electrostatic potential energy within the COFs pores\cite{zojer2023electrostatic,zojer2024tuning}, which have been found to facilitate charge separation in experiments\cite{COF2014Medina}. While a number of organic molecules have been found to form van der Waals heterostructures with TMD\cite{Asterdam2019ACSNano,zhou2021range} and Janus TMD\cite{Prezhdo2024JPCL} 2D materials, organic-nanotube van der Waals heterostructures are expected to form with those molecules and Janus TMD nanotubes of sufficient pore sizes. The energy level alignment tuned by the electrostatic potential can significantly modify charge transfer and catalytic efficiencies. In experiments, the energy level alignment can be obtained through scanning tunneling spectroscopy (STS)\cite{leroy2004scanning, carroll1998local}, which provides a direct access to the local density of states.

By changing the direction and magnitude of the radially distributed dipoles determined by the relative electronegativity of the surface atoms of the nanotube, the electrostatic potential inside the nanotube can be modified. Along with the Janus nanotube database\cite{bolle2021structural}, the analytical model in this work can be used to identify and design 1D Janus nanotubes and their heterostructures for applications in optoelectronics, %molecule storage and separation, energy storage, 
and (photo)catalysis.

 In summary, Janus TMD nanotubes generate a large and uniform electrostatic potential at the interior of the tube, which increases in magnitude as the nanotube radius increases, approaching that for the corresponding 2D sheets. We developed an analytical formula for the electrostatic potential, which shows its strong dependence on the quadrupole moment and size of the nanotube. We show that these electrostatic effects can be described by representing the charge density of the nanotube as radially distributed dipoles, arranged periodically in 1D along the tube axis. For DW nanotubes, the interior electrostatic potential is accumulative of the constituent SW nanotubes, though reduced by interface charge redistribution. We find that these electrostatic effects of the nanotube can tune the band alignment for nanotube heterostructures. In a DW nanotube, the band energies of the inner tube are significantly shifted by about 1.0 eV, which is originated from the electrostatic potential of the outer tube. Dielectric screening effects slightly reduce the band gaps of both the inner and outer tubes, on the order of magnitude of 0.1 eV. The electrostatic potential within the pores of the nanotube is largely tunable by changing the radial polarization, which provides a versatile strategy to tune the electronic and optical properties of nanomaterials. % which is promising for various applications. 

%  quadrupole moment model to describe the electrostatic potential in Janus nanotubes. For single-walled nanotubes, the effective charge and quadrupole moment increase with diameter, leading to a higher potential shift for larger tubes. Further, electronic calculations of the double-wall suggest a significant change for the electronic states of the inner tube, while the electronic states of the outer tube changed marginally. The observation can be explained by considering the electrostatic effects - interlayer polarization and dielectric screening. Our study emphasizes the role of electrostatics for the better understanding of band alignment in Janus nanotubes. However, our approach overlooks the finer details, such as orbital hybridization and spin-orbit coupling effects for band alignment. Even though, PBE qualitatively predicts the electrostatic trends reliably, the quantitative description of the band edge position should be verified by using hybrid functionals or by accounting for GW-level corrections. Additionally, we considered n=6 and n=14 combinations of nanotubes to investigate the band alignment in the double wall. In nanotubes, chirality, strain, and rotation could cause additional screening effects in the materials. \par    

%%%%%%%%%%%%%%%%%%%%%%%%%%%%%%%%%%%%%%%%%%%%%%%%%%%%%%%%%%%%%%%%%%%%%
%% The "Acknowledgement" section can be given in all manuscript
%% classes.  This should be given within the "acknowledgement"
%% environment, which will make the correct section or running title.
%%%%%%%%%%%%%%%%%%%%%%%%%%%%%%%%%%%%%%%%%%%%%%%%%%%%%%%%%%%%%%%%%%%%%
\begin{acknowledgement}

This work is primarily supported by US National Science Foundation (NSF), EPSCoR, Grant No. OIA-2521414, and the University of Kansas New Faculty Research Development Award and Undergraduate Research Award. R.S. acknowledges the support from the NSF REU program at the University of Kansas with award No. PHY-2447841. Computational resources are from Argonne National Laboratory and the Pittsburgh Supercomputing Center. Work performed at the Center for Nanoscale Materials, a U.S. Department of Energy Office of Science User Facility, was supported by the U.S. DOE, Office of Basic Energy Sciences, under Contract No. DE-AC02-06CH11357. This work used Bridges-2 at the Pittsburgh Supercomputing Center through allocation PHY230195 from the Advanced Cyberinfrastructure Coordination Ecosystem: Services \& Support (ACCESS) program, which is supported by the U.S. National Science Foundation, Grant Nos. 2138259, 2138286, 2138307, 2137603, and 2138296.

\end{acknowledgement}

%%%%%%%%%%%%%%%%%%%%%%%%%%%%%%%%%%%%%%%%%%%%%%%%%%%%%%%%%%%%%%%%%%%%%
%% The same is true for Supporting Information, which should use the
%% suppinfo environment.
%%%%%%%%%%%%%%%%%%%%%%%%%%%%%%%%%%%%%%%%%%%%%%%%%%%%%%%%%%%%%%%%%%%%%
\begin{suppinfo}
\begin{itemize}
\item Computational details; The mathematical derivation of the 1D periodic electrostatic potential, Eq.1 and Eq. 2; calculation of effective charges; convergence test of $n=6$ nanotube; The in-plane electrostatic potential of the investigated MoSTe nanotubes and double-wall nanotube; z-dependence of electrostatic potential; charge density redistribution for the double-wall MoSTe nanotube; DOS and x,y-averaged electrostatic potential of inner and outer tubes of the DW MoSTe nanotubes;The in-plane electrostatic potential of MoS$_2$ SW and DW; The TDOS of SW and DW MoS$_2$ and band alignment diagram of DW MoS$_2$;The electrostatic potential of zigzag $n=10$ MoSTe nanotube; numerical details of calculated potential and quadrupole using Eq.1 and Eq.2.
\item  All structures used in this study (ZIP).
\end{itemize}
\end{suppinfo}

%\section*{Data Availability Statement}
%The data underlying this study are openly available on Zenodo.org at \\
%https://doi.org/10.5281/zenodo.19323296.

%%%%%%%%%%%%%%%%%%%%%%%%%%%%%%%%%%%%%%%%%%%%%%%%%%%%%%%%%%%%%%%%%%%%%
%% The appropriate \bibliography command should be placed here.
%% Notice that the class file automatically sets \bibliographystyle
%% and also names the section correctly.
%%%%%%%%%%%%%%%%%%%%%%%%%%%%%%%%%%%%%%%%%%%%%%%%%%%%%%%%%%%%%%%%%%%%%
%\bibliographystyle{chem-acs}
\bibliography{nanotube}
\end{document}

\maketitle
\section{Computational Details}
\renewcommand{\thefigure}{S\arabic{figure}}
\renewcommand{\theequation}{S\arabic{equation}} %and \setcounter{equation}{0} 

The electrostatic potential of the nanotube was calculated using the PWscf package in Quantum ESPRESSO \cite{giannozzi2009quantum}, with Perdew–Burke–Ernzerhof (PBE) exchange–correlation functional within the generalized gradient approximation (GGA) \cite{perdew1996generalized} . The Optimized Norm-Conserving Vanderbilt Pseudopotential (ONCVPSP) obtained from the Pseudo-Dojo library\cite{Hamann2018pseudodojo} was used for potential description \cite{hamann2013optimized}. The ecutwfc of 90 Ry and K-point mesh of \(1 \times 1 \times 8\) was used for structural optimization and self-consistent field (SCF) calculations. The energy convergence criteria were set to  \(10^{-8}\) for electronic calculations. \par  
The density of states (DOS) and projected DOS (PDOS) analysis were made using Vienna Ab initio Simulation Package (VASP) \cite{kresse1996efficient, kresse1993ab} . The projector augmented wave (PAW) method within the PBE was used for electron-ion interaction \cite{holzwarth1997comparison, kresse1999ultrasoft} . DFT-D3 correction was applied to account for the vdW corrections in the double-wall MoSTe nanotube \cite{grimme2011effect} . The plane wave energy cut-off of 550 eV and electronic convergence criteria of \(10^{-6}\) were used for SCF calculations. %A denser K point mesh of \(1 \times 1 \times 24\) was used for DOS calculations. 
A k-point mesh of \(1 \times 1 \times 8\) and \(1 \times 1 \times 24\) was used for SCF and DOS calculations, respectively. The post-processing analysis is performed using pymatgen \cite{ong2013python} . For all investigated structures, a vacuum of 21 \AA{} was applied along $x$ and $y$ directions to avoid any interactions resulting from periodic images.  

\begin{figure} [H]
	\centering
	\includegraphics[width=0.95\linewidth]{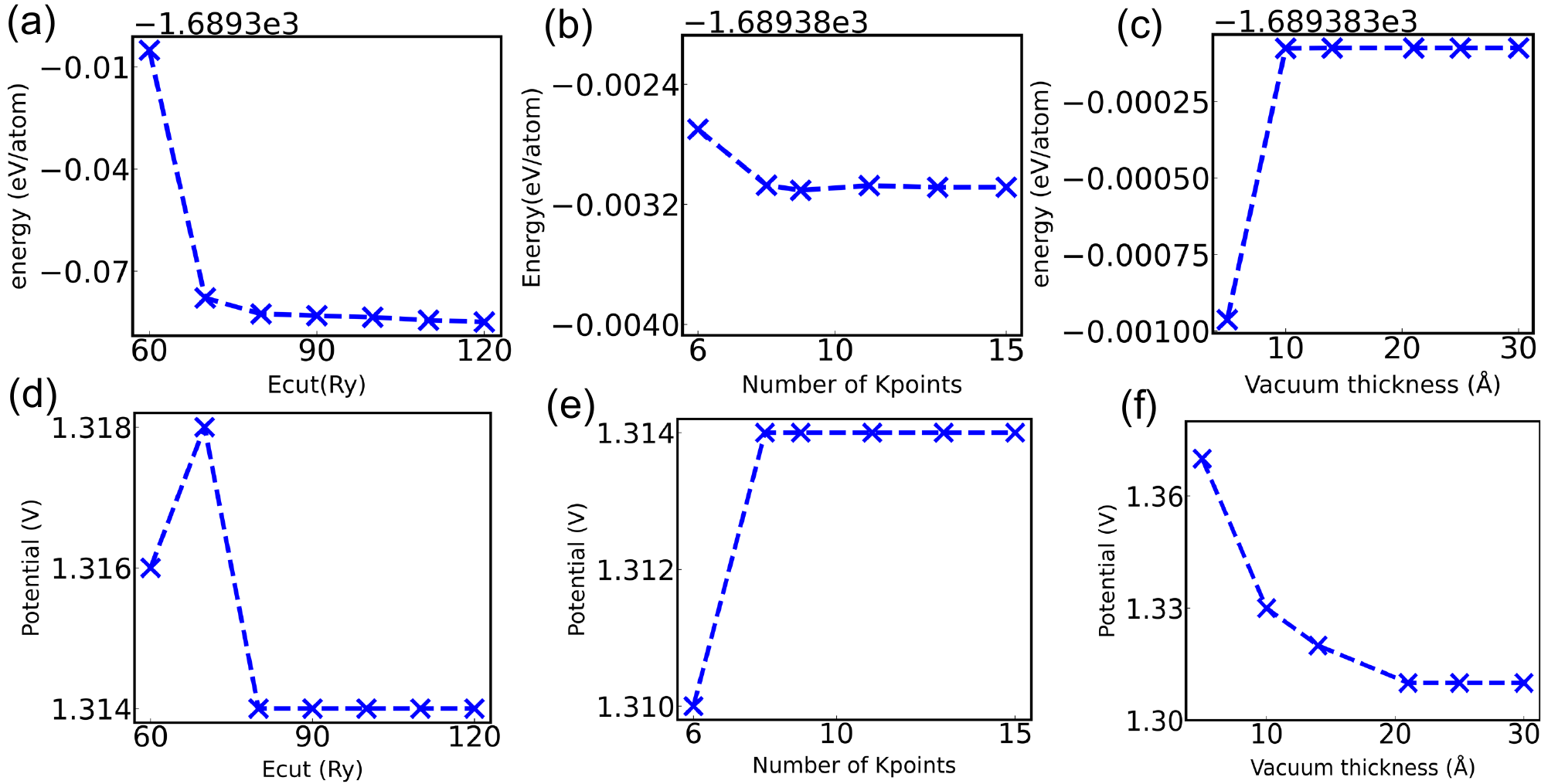}
	\caption{Convergence test with Quantum ESPRESSO for total energy and electrostatic potential at the center of the $n=6$ SW nanotube as a function of (a, d) plane-wave energy cutoff, (b,e) number of k-points along $z$ direction, and (c,f) lateral vacuum thickness. 
			%    (a) k-point convergence of the total energy for the $n = 6$ MoSTe nanotube using a $1 \times 1 \times N$ mesh.
			%    (b) Convergence of the total energy with respect to the plane-wave cutoff energy.
			%    (c) Convergence of the total energy with respect to the lateral vacuum thickness.
	}
	\label{fig:S1}
\end{figure}

\begin{figure} [H]
	\centering
	\includegraphics[width=0.68\linewidth]{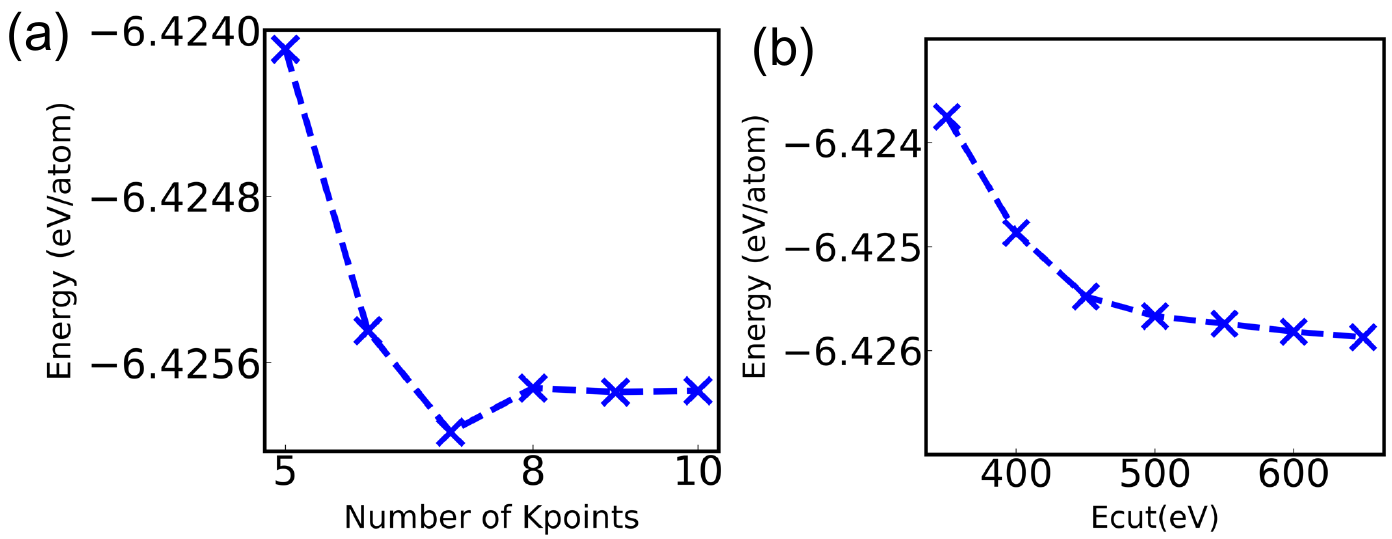}
	\caption{Convergence test on (a) number of k points along the $z$ direction, (b)  plane-wave energy cutoff, E$_{cut}$, using the $n=6$ SW MoSTe nanotube. Calculations were performed using VASP.% for  and the convergence test on (c) total energy, and (d) interior potential, for lateral vacuum thickness.
			%    (a) k-point convergence of the total energy for the $n = 6$ MoSTe nanotube using a $1 \times 1 \times N$ mesh.
			%    (b) Convergence of the total energy with respect to the plane-wave cutoff energy.
			%    (c) Convergence of the total energy with respect to the lateral vacuum thickness.
	}
	\label{fig:S1-1}
\end{figure}

The plane-wave energy cutoff, number of k-points, and lateral vacuum thickness were chosen to achieve convergence of 1 meV/atom for total energy, and 1 mV for potential, are shown in Fig.~\ref{fig:S1} and Fig.~\ref{fig:S1-1}. All atomic structure are visualized and image generated using VESTA\cite{momma2008vesta}.

\begin{figure}[H]
    \centering
    \includegraphics[width=0.75\linewidth]{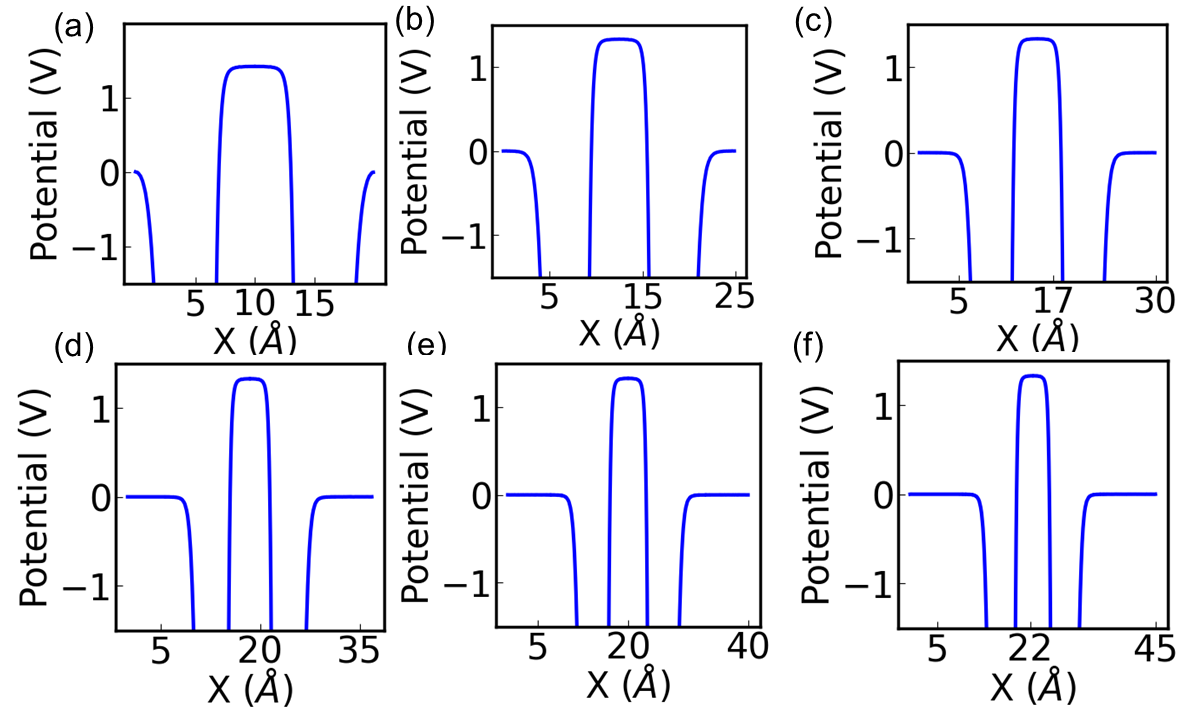}
    \caption{The real space electrostatic potential of the $n=6$ MoSTe nanotube, at $Z=1 $ \AA\ plane, along the $y=0$ \AA\ line as a function of lateral vacuum size, (a) 5 \AA, (b) 10 \AA, (c) 14 \AA, (d) 21 \AA, (e) 24 \AA, (f) 30 \AA. %(d) convergence of the potential at the center of the nanotube as a function of lateral vacuum size
    }
    \label{fig:S2}
\end{figure}

\section{Electrostatic potential for MoSTe Janus nanotubes}

\begin{figure}[H]
    \centering
    \includegraphics[width=0.98\linewidth]{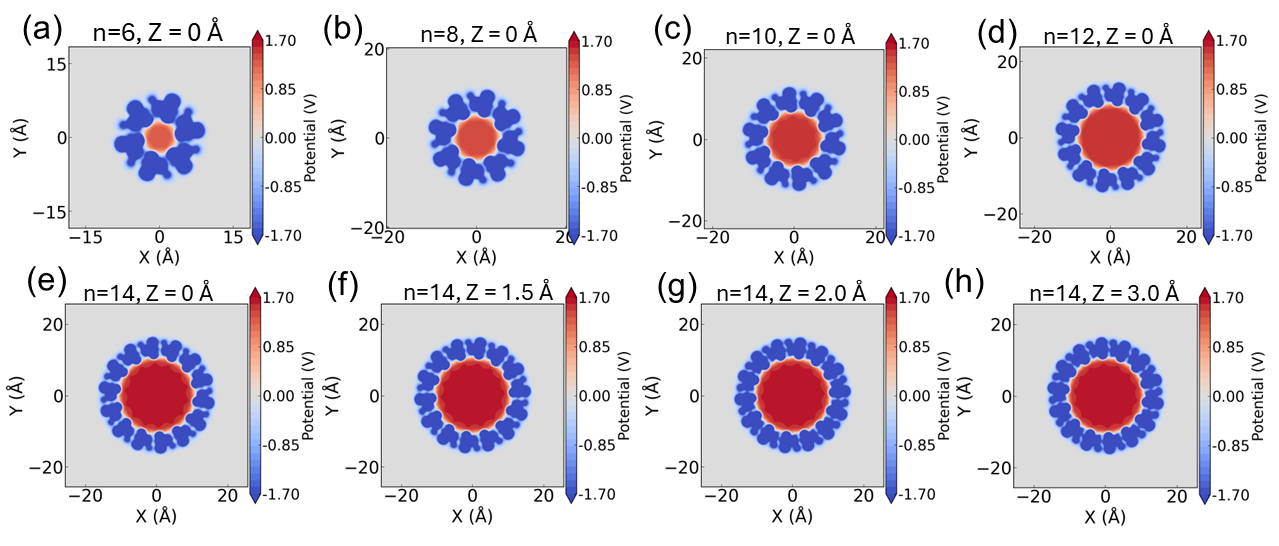}
    \caption{(a) to (d) Real space electrostatic potential of nanotubes with $n = 6,8,10$ and $12$ respectively, in the $Z = 0$ \AA{} plane(e) to (h) Real space electrostatic potential of $n = 14$ nanotube in the $Z = 0$ \AA{}, $Z = 1.5$ \AA{}, $Z = 2$ \AA{} and $Z = 3.0$ \AA{} planes respectively. All potentials are referenced to the vacuum potential.}
    \label{fig:S3}
\end{figure}
\begin{figure}
    \centering
    \includegraphics[width=0.75\linewidth]{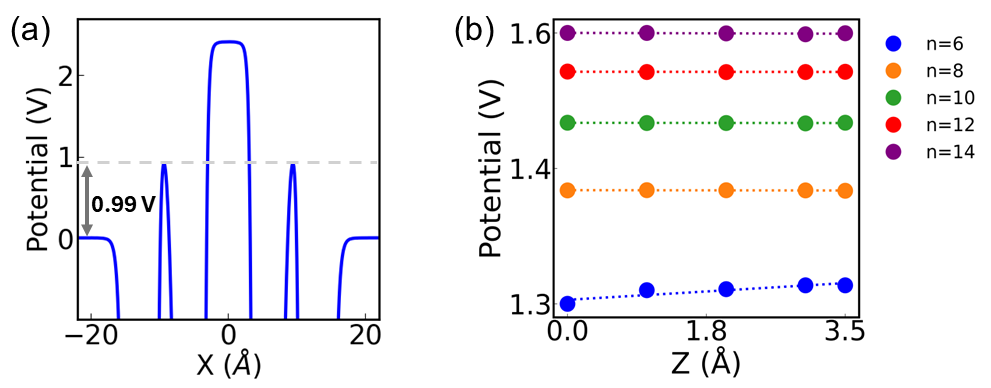}
    \caption{(a) Real space electrostatic potential of the DW MoSTe along the $y = 0$ \AA{} line; the dashed line indicates the potential at the interface due to the charge redistribution (b) Real space electrostatic potential as a function of $z$ coordinates for different radius nanotubes. All potentials are referenced to vacuum.}
    \label{fig:S4}
\end{figure}
\begin{figure}
    \centering
    \includegraphics[width=0.75\linewidth]{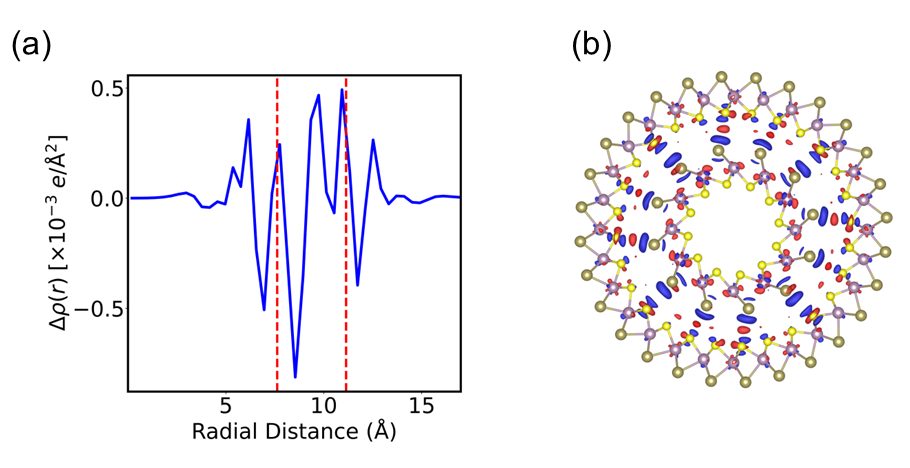}
    \caption{(a) The charge density difference, $\Delta\rho = \rho_{\mathrm{DW}} - \rho_{\mathrm{inner}} - \rho_{\mathrm{outer}}$ of the DW MoSTe along the radial direction. The red lines indicate the position of the inner Te and outer S atoms, respectively. (b) The visualization of $\Delta\rho$, with electron accumulation (red) and electron depletion (blue).}
    \label{fig:S5}
\end{figure}
\begin{figure}
    \centering
    \includegraphics[width=0.5\linewidth]{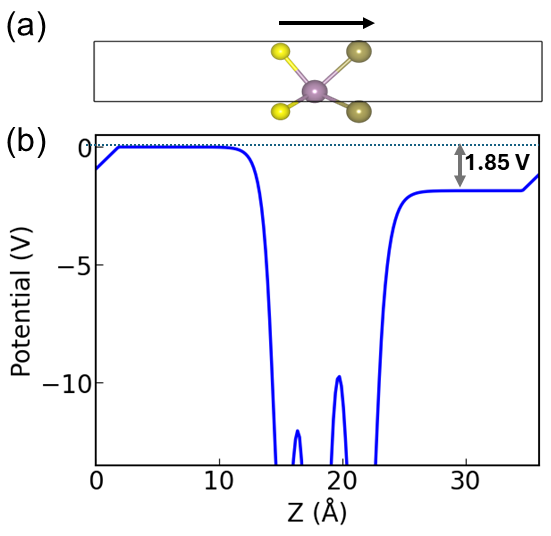}
    \caption{(a) Atomic structure of 2D MoSTe monolayer, with arrow indicating the out-of-plane polarization (b) z-averaged electrostatic potential showing the potential drop of 1.85 V across the layer. The potential is referenced to vacuum.}
    \label{fig:S6}
\end{figure}
\begin{figure}
    \centering
    \includegraphics[width=0.8\linewidth]{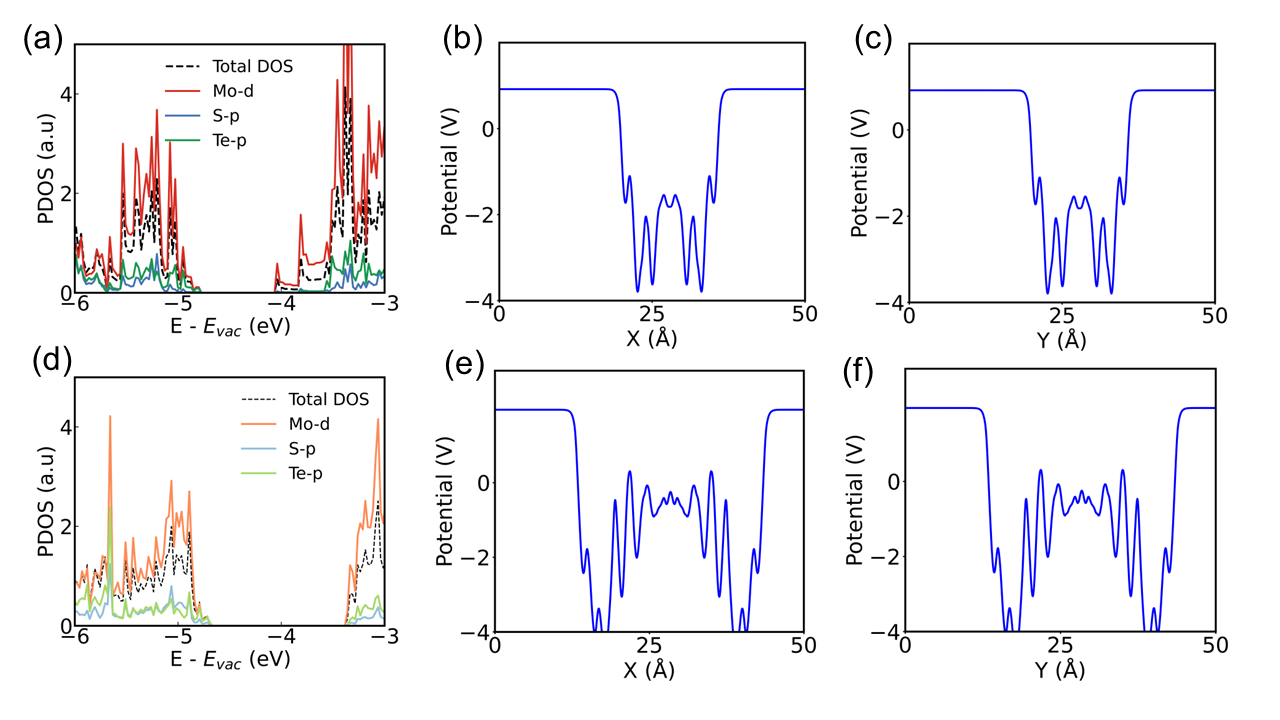}
    \caption{(a) Normalized projected density of states (pDOS) of Mo, Te, and S atoms in the inner MoSTe tube of the DW. The black dashed line shows the normalized total density of states (DOS) of the inner tube. The energies are referenced to the vacuum  (b) $y$ -averaged electrostatic potential of the inner tube. (c) $x$-averaged electrostatic potential of the inner tube. (d) Normalized pDOS of Mo, Te, and S atoms of the outer tube of the DW, with the black dashed line showing the normalized total DOS of the outer tube. The energies are referenced to the vacuum  (e)  $y$-averaged electrostatic potential of the outer tube (f) $x$-averaged electrostatic potential of the outer tube.}
    \label{fig:S7}
\end{figure} 

\begin{figure}
	\centering
	\includegraphics[width=0.65\linewidth]{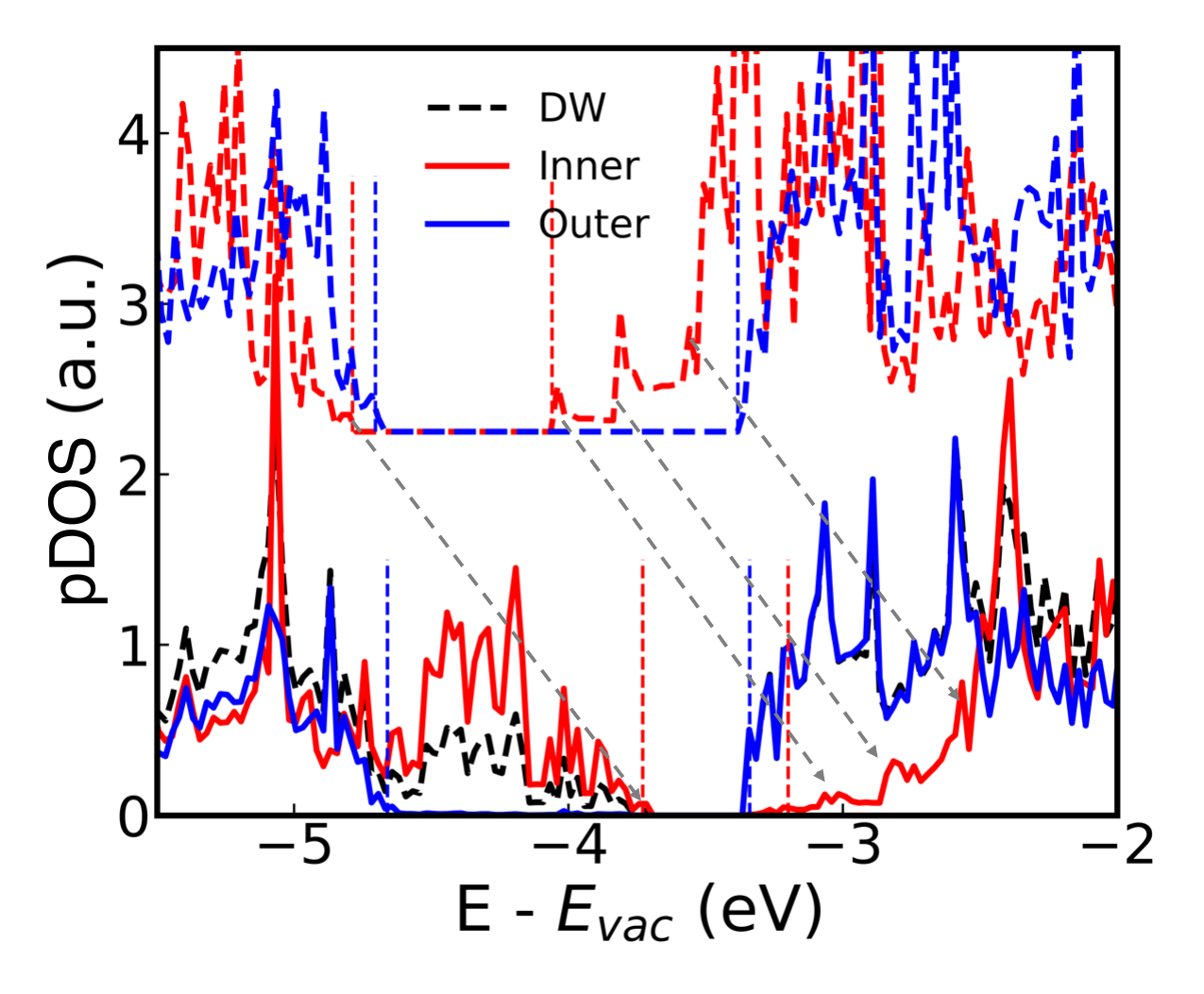}
	\caption{The normalized total density of states (tDOS) of the DW MoSTe nanotube (black dashed line), and projected DOS (pDOS) on the inner tube (solid red line) and outer tube (solid blue line). The normalized tDOS of the corresponding isolated SW $n=6$ (red dashed line) and $n=14$ (blue dashed line) nanotubes. The vertical dashed lines correspond to the VBM/CBM of the $n=6$ (red dashed line), and $n=14$ (blue dashed lines) in the SW and DW nanotubes. The dashed gray lines connects the intrinsic states for the $n=6$ tube in the SW and DW nanotubes, showing a consistent shift of about 1.05 eV in energies. The CBM (dashed red vertical line) of the inner tube within the DW MoSTe is a result of interface-induced electronic hybridization with the conduction bands of the outer tube, which leads to a band gap reduction for the inner tube.}
	\label{fig:SIpDOSMoSTe}
\end{figure} 

\begin{figure}
    \centering
    \includegraphics[width=0.75\linewidth]{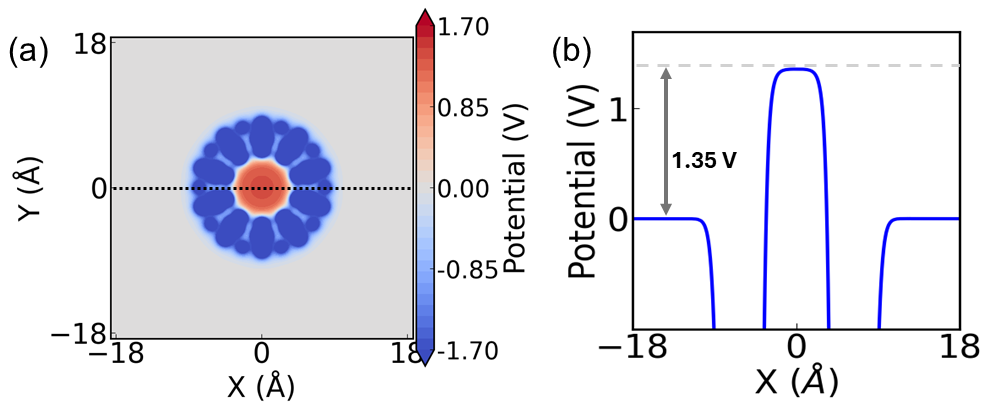}
    \caption{(a) The real-space electrostatic potential of the $n=10$ zigzag MoSTe nanotube at the $z=0.7$\AA\ plane, and (b) Electrostatic potential of the nanotube along the $y=0$ \AA\ line marked in (a).
    The magnitude of the uniform interior potential for $n=10$ Zigzag MoSTe, which has a radius of about 8.8\AA, is similar to that for $n=6$ armchair MoSTe nanotube, which has a slightly smaller nanotube radius of about 8.6 \AA. Meanwhile, the interior electrostatic potential for the zigzag MoSTe from Eq. 2 agrees well with that from DFT, see Table S1. This demonstrates that the electrostatic effects and analytical formula derived based on the armchair MoSTe nanotube is transferable to 1D nanotubes of different chiralities.}
    \label{fig:S10}
\end{figure}

\begin{table}[htbp]
	\centering
	\caption{The effective charge $q$ $(e)$ value calculated using Eq. \ref{eq:q_calcu}, the corresponding quadrupole moment calculated using Eq.~\ref{eq:quadrupole_1}, $Q$ ($e$\AA$^2$), the  geometrical factor $R_{Te}^2-R_s^2$ (\AA$^2$), thickness $R_{Te}-R_s$ (\AA), $Q'$ ($e$\AA$^2$) calculated using the same $q$ value as 2D MoSTe, $q=0.0298e$. The effective charge $q_c$ $(e)$ calculated using Eq.~\ref{eq:V_Colu}. Electrostatic potential inside the MoSTe nanotube from DFT ($V_{DFT}$), from Eq. 1 with the corresponding $q$ ($V_{eq1}$), from Eq. 2 with the corresponding $q$ ($V_{eq2}$), and from Eq. 2 with the same $q$ value of  2D MoSTe ($V'_{eq2}$). The units for the electrostatic potentials are in volt (V). For the DW, first and second values correspond to the inner and outer tubes values for the $R_{Te}^2-R_S^2$ and $R_{Te}-R_S$ respectively. Data for $n=10$ zigzag MoSTe nanotube is also included.}%The quadrupole and potential values reported for the DW as a whole.}
%\textcolor{red}{put the table in the SI?}}
\label{tab:S1}
\begin{tabular}{@{}cccccccccccc@{}}
\toprule
%        \makecell{MoSTe \\ size (n)} &
%        \makecell{Effective \\ charge (e)} &
%        \makecell{Quadrupole \\ moment (e$\text{\AA}^{2}$)} &
%        \makecell{Potential \\ V\_DFT (V)} &
%        \makecell{Potential \\ V\_q\_FIT (V)} &
%        \makecell{Potential \\ V\_q\_COM (V)} \\
$n$ & $q$ & $q_c$ & $R_{Te}^2-R_S^2$ & $R_{Te}-R_S$ & $Q$ & $Q'$ & $V_{DFT}$ & $V_{eq1}$ & $V_{eq2}$ & $V'_{eq2}$ \\
\midrule
6  & 0.0230 & 0.0231 & 39.20 & 3.27 & 10.84 & 14.02 & 1.31 & 1.31 & 1.31 &  1.69\\
8  & 0.0239 & 0.0239 & 51.33 & 3.33 & 19.64 & 24.48 & 1.42 & 1.42 & 1.42 & 1.77 \\
10 & 0.0245 & 0.0245 & 63.62 & 3.36 & 31.20 & 37.91 & 1.49 & 1.49 & 1.49 & 1.82 \\
12 & 0.0250 & 0.0250 & 75.97 & 3.37 & 45.73 & 54.33 & 1.55 & 1.55 & 1.55 & 1.84 \\
14 & 0.0255 & 0.0256 & 88.35 & 3.38 & 63.34 & 73.72 & 1.60 & 1.60 & 1.60 & 1.85 \\
\multirow{2}{*}{DW} & \multirow{2}{*}{0.0198} & \multirow{2}{*}{0.0198} & 39.20, & 3.29,  & \multirow{2}{*}{57.86} & \multirow{2}{*}{86.95} & \multirow{2}{*}{2.41} & \multirow{2}{*}{2.41} & \multirow{2}{*}{2.41} & \multirow{2}{*}{3.62} \\
& & &  87.71  & 3.40 & & & &  & & \\
%DW & 0.0198 & 0.0198 & 39.20, 87.71 & 3.29, 3.40 & 57.86 & 86.95 & 2.41 & 2.41 & 2.41 & 3.62 \\
%           &       &       & (in,out) & (in,out) &  &  &  &  &  &    \\ 
%\textcolor{blue}{10}&\textcolor{blue}{0.0236}&\textcolor{blue}{0.0236}&\textcolor{blue}{39.95}& \textcolor{blue}{3.26}& \textcolor{blue}{18.86}& \textcolor{blue}{23.81}& \textcolor{blue}{1.35}& \textcolor{blue}{1.35}& \textcolor{blue}{1.35}& \textcolor{blue}{1.70}\\
10&\multirow{2}{*}{0.0236}&\multirow{2}{*}{0.0236}&\multirow{2}{*}{39.95}& \multirow{2}{*}{3.26}& \multirow{2}{*}{18.86}& \multirow{2}{*}{23.81}& \multirow{2}{*}{1.35}& \multirow{2}{*}{1.35}& \multirow{2}{*}{1.35}& \multirow{2}{*}{1.70}\\
(zigzag) & & & & & & & & & & \\
\bottomrule
\end{tabular}
\end{table}

\vspace{10pt}
\textcolor{blue}{\section{Electrostatic potential and band alignment for MoS$_2$ nanotubes}}
\begin{figure}
	\centering
	\includegraphics[width=0.95\linewidth]{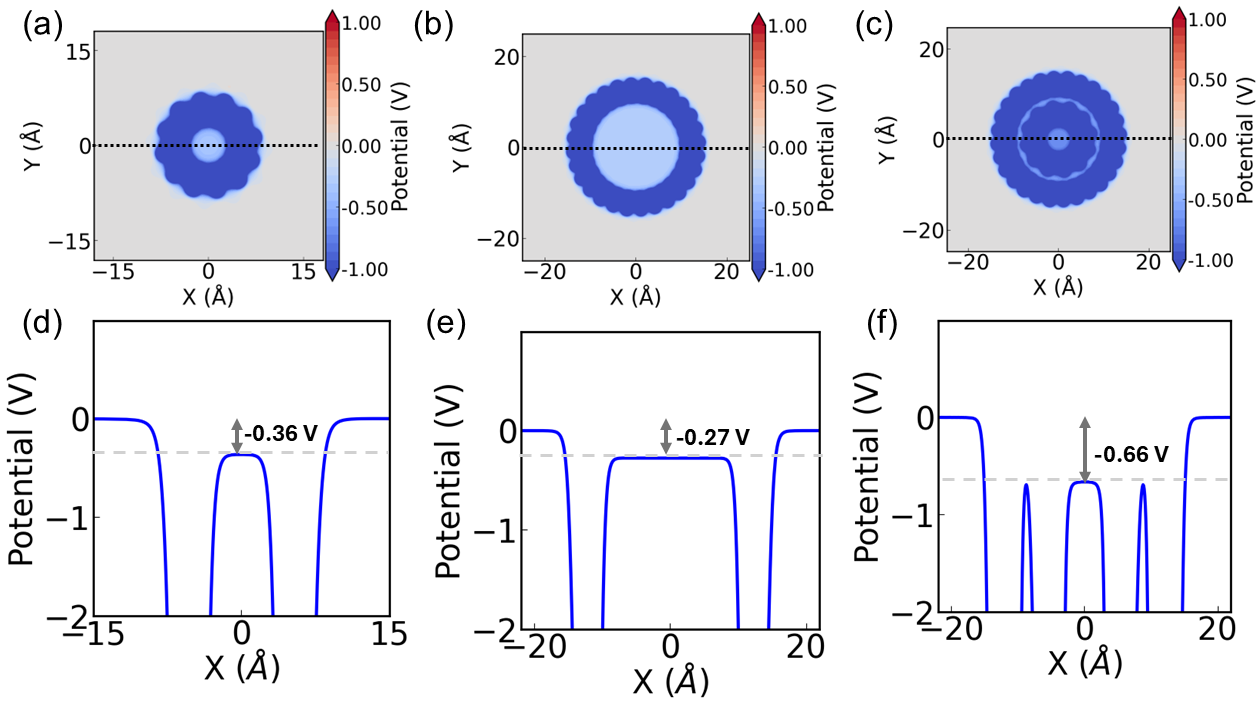}
	\caption{The real-space electrostatic potential at the $z=1.0$ \AA\ plane for (a) SW MoS$_2$ nanotube ($n = 6$), (b) SW MoS$_2$ ($n=14$) (c) DW MoS$_2$ nanotube consisting of the $n=6$ and $n=14$ tubes. Electrostatic potential along the $y = 0$ \AA\ direction for (d) SW MoS$_2$ ($n = 6$), (e) SW MoS$_2$ ($n=14$) and (f) DW MoS$_2$ nanotube S. All potentials are referenced to the vacuum potential. $z$-axis is along the tube axis.}
	\label{fig:S8}
\end{figure}

\begin{figure}
	\centering
	\includegraphics[width=0.95\linewidth]{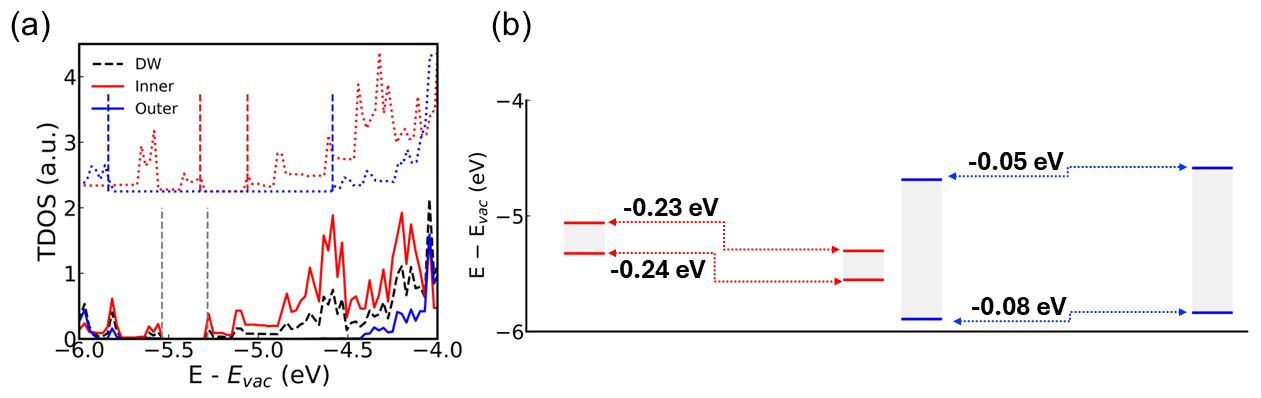}
	\caption{(a) The normalized total density of states (TDOS) of the DW MoS$_2$ nanotube (black dashed line), with contributions from the inner tube (solid red line) and outer tube (solid blue line). The normalized TDOS of the corresponding isolated SW nanotubes are shown for comparison: inner wall (red dotted line) and outer wall (blue dotted line). The vertical dashed lines correspond to the VBM/CBM of the DW (grey dashed line), SW $n=6$ (red dashed line), SW $n=14$ (blue dashed lines). (b) VBM and CBM band energies for the $n=6$ (red) and $n=14$ (blue) SW MoS$_2$ nanotubes and that in the DW MoS$_2$nanotube. All energies are referenced to the vacuum. }
	\label{fig:S9}
\end{figure}

For comparison, we also study the electrostatic potential (Fig.~\ref{fig:S8}) and electronic states (Fig.~\ref{fig:S9}) of the $n=6$ and $n=14$ SW MoS$_2$ and the corresponding DW MoS$_2$ formed by combining $n=6$ and $n=14$ nanotubes. The MoS$_2$ nanotubes also generates uniform electrostatic potentials at the nanotube interior, -0.36 V and -0.27 V for the $n=6$ and $n=14$ SW nanotubes, respectively. %In contrast to Janus MoSTe, MoS$_2$ lacks intrinsic polarization at the 2D limit, and the electrostatic potential solely arises from the curvature induced charge redistribution %, i.e, flexoelectric effect in TMD nanotubes
%\cite{zhao2023curvature, dong2021spontaneous}. 
The the magnitude is much smaller than for MoSTe nanotubes of the same radii. As there's no intrinsic radial polarization for MoS$_2$ nanotubes, this potential inside the tube is solely resulted from the curvature-induced charge redistribution––less accumulation of electron densities at the inner S atoms, leading to relatively more positive charge densities than for the outer S atoms. Therefore the electrostatic potential per electron (unit of negative charge) at the nanotube interior is negative, in contrast to that for MoSTe nanotubes. 
%As a result, the interior of the  MoS$_2$ nanotubes shows a negative potential with a magnitude of the potential larger for smaller radius nanotubes. 
\\
For the DW MoS$_2$, the potential at the center of the nanotubes, -0.66 V, is close to the sum of potential from the $n=6$, -0.36 V, and $n=14$, -0.27 V, with a deviation of about 23 mV, suggesting weak interfacial charge transfer due to the same S atoms at the interface for both tubes. From Fig.~\ref{fig:S9}(a), the electronic states of the individual SW MoS$_2$ are preserved in DW MoS$_2$, with a band energy VBM(CBM) shift of about -0.24 V (-0.23 V) for the inner tube comparing with the SW $n=6$ tube. This band energy shift is similar to the interior potential of the $n=14$ outer tube. In contrast, the changes in VBM and CBM for the outer tube are negligible, since the electrostatic potential outside of the inner tube is zero. The band gap of the inner tube is slightly reduced from 0.25 eV to 0.23 eV, while the band gap of the outer tube increases moderately from 1.25 eV to 1.38 eV. We note that the extraction of the band edge from the TDOS of the DW involves a finite threshold states, which can slightly influence the identification of low weight states close to the band edges. Nevertheless, the overall trend of band edge shifts compared to the SW nanotube remains robust.

\FloatBarrier
\section{Electrostatic potential derivation}
The electrostatic potential at a point $(x,y,z)$ due to a 1D periodic arrangement of point charges along the $z$-direction is given by Coulomb’s law,
\begin{equation}
    V = K \sum_{j,n} \frac{q_j}{\sqrt{(x-x_j)^2+(y-y_j)^2+(z-z_j-nc)^2}} 
    \label{eq:V_Colu}
\end{equation}
where $K = 14.39~\mathrm{eV\,\text{\AA}^{}/e^2}$ is the Coulomb constant, $c$ is the lattice constant along the nanotube axis. The position of charge $q_j$, $(x_j, y_j, z_j)$ is located at the positions of S and Te atoms in the DFT-relaxed structures. $n$ is the number of periodic unit cells along the nanotube axis.
The equation can be rewritten using the dimensionless terms as, 
\begin{equation}    
V =  \frac{K}{c} \sum_j q_j \sum_n \frac{1}{\sqrt{\alpha^2+\beta^2+(\delta-n)^2}} 
\end{equation}
with
\begin{equation}
    \alpha = \frac{x-x_j}{c},
     \beta = \frac{y-y_j}{c},
      \delta = \frac{z-z_j}{c}
\end{equation}
For one-dimensional periodic systems, the Epstein-Hurwitz $\zeta$ function is given by,
\begin{equation}
   \zeta_{1D}(s=\frac{1}{2}) = \sum_n \frac{1}{\sqrt{(\alpha^2+\beta^2+(\delta-n)^2}}
   \label{eq:Eps_Hur}
\end{equation}
Using the Eq~.(\ref{eq:Eps_Hur}), the potential can be expressed as
\begin{equation}
    V = \frac{K}{c} \sum_j q_j \zeta_{1D}(s= \frac{1}{2})
    \label{eq:pot_zeta}
\end{equation}
Mellin transform of a function, $\theta(t)$ is defined as, 
\begin{equation}
    \tilde{\theta}(s) = \int_0^\infty{t^{s-1}\theta(t) dt}
\end{equation}
The Mellin transform is related to the Epstein-Hurwitz $\zeta$ function \cite{SI_zhou2021engineering}, 
\begin{equation}
    \tilde{\theta}(s) = \pi^{-s} \Gamma(s)\zeta(s)
    \label{eq:ep_zeta}
    \end{equation}
We define a $\theta$-function, whose Mellin transform is related to the Epstein-Hurwitz  $\zeta $ function,  
\begin{equation}
    \theta(t) = \sum_n \exp{(-\pi(\alpha^2+\beta^2+(\delta-n)^2)t}
    \label{eq:zeta_theta}
\end{equation}
Using Poisson summation and Poisson-Gaussian identity,

   % \sum_n \exp{(-\pi(\delta-n)^2}t) = \frac{1}{\sqrt{t}}\sum_k \exp{(-2\pi i k\delta)\exp(\frac{-\pi k^2}{t})} 
   \begin{equation}
\sum_{n=-\infty}^{\infty}
\exp\!\left(-\pi(\delta-n)^2 t\right)
=
\frac{1}{\sqrt{t}}
\sum_{k=-\infty}^{\infty}
\exp\!\left(2\pi i k \delta\right)
\exp\!\left(-\frac{\pi k^2}{t}\right)
\label{eq:theta_poisson}
\end{equation}
Using Eq.~(\ref{eq:theta_poisson}), Eq.~(\ref{eq:zeta_theta}) can be rewritten as, 

\begin{equation}
\theta(t) =
\frac{e^{-\pi(\alpha^2+\beta^2)t}}{\sqrt{t}}
\sum_{k=-\infty}^{\infty}
\exp\!\left(2\pi i k \delta\right)
\exp\!\left(-\frac{\pi k^2}{t}\right)
\label{eq:theta_k}
\end{equation}
at $k=0$,
\begin{equation}
\theta_{k=0}(t) =
\frac{e^{-\pi(\alpha^2+\beta^2)t}}{\sqrt{t}}
\label{eq:theta_k_0}
\end{equation}
For $k=0$, the Mellin transform diverges at the lower limit $t \to 0$. To handle this, we introduce a small cutoff $\epsilon > 0$ for $k=0$ 
\begin{equation}
\tilde{\theta} =
\sum_{k\neq0}
e^{2\pi i k \delta}
\int_0^\infty
\frac{dt}{t}
\exp\!\left(-\pi(\alpha^2+\beta^2)t-\frac{\pi k^2}{t}\right)
\label{eq:theta_mellin}
\end{equation}

for $k=0$ term, 

\begin{equation}
\int_{\epsilon}^{\infty}
\frac{dt}{t}
e^{-\pi(\alpha^2+\beta^2)t}
=
-\gamma
-2\ln\!\left(\epsilon\sqrt{\alpha^2+\beta^2}\right)
+\mathcal{O}(\epsilon)
\label{eq:v_part_k_0}
\end{equation}
with $\gamma$ is the Euler–Mascheroni constant.
The modified Bessel function of the second kind is defined as

\begin{equation}
K_\nu(2\sqrt{ab}) =
\frac{1}{2}\left(\frac{a}{b}\right)^{\nu/2}
\int_0^\infty t^{\nu-1} e^{-a/t - b t}\,dt.
\label{eq:bessel_fun}
\end{equation}

Using $\nu = 0$, $a = \pi k^2$, and $b = \pi(\alpha^2+\beta^2)$, Eq.~(\ref{eq:theta_mellin}) can be rewritten interms of  Eq.~(\ref{eq:bessel_fun}) as, 
\begin{equation}
\tilde{\theta}
=
\sum_{k \neq 0}
\exp{(-2\pi i k \delta)}
\kappa_{0}\!\left(2\pi |k|\sqrt{\alpha^2+\beta^2}\right)
\label{eq:theta_Modi_Bess_correct}
\end{equation}

Using Eq.~(\ref{eq:theta_Modi_Bess_correct}), Eq.~(\ref{eq:ep_zeta}) can be rewritten as,
\begin{equation}
    \zeta(s=\frac{1}{2}) = \frac{2}{\pi} \sum_{k \neq 0} \exp{(-2\pi i k \delta) \kappa_0\left(2\pi |k| \sqrt{(\alpha^2+\beta^2)}\right)}
    \label{eq:zeta_bessel}
\end{equation}

substituting Eq.~(\ref{eq:v_part_k_0}), and Eq.~(\ref{eq:zeta_bessel}) to 
Eq.~(\ref{eq:pot_zeta}), 
\begin{equation}
V = \frac{2K}{c} 
\sum_j q_j \left(
-\ln R_j
+ \sum_{k \neq 0} \frac{1}{\pi}
\exp(-2\pi i k \delta)\,
\kappa_0\!\left(2\pi |k| \frac{R_j}{c}\right)
\right)
\label{eq:V_final}
\end{equation}
with 
\begin{equation}
    R_j = \sqrt{(x-x_j)^2+(y-y_j)^2}
\end{equation}
where \( q_j \) are the effective point charges, \( R_j \) their distance from the subject point ($x,y$). The constant term, $-\gamma - 2\ln \epsilon$ for the $k=0$ term is independent of the in-plane coordinates and 
does not affect potential differences and is therefore excluded from the final expression. 

\subsection{Electrostatic potential in terms of quadrupole moment}
Assuming equal and opposite charges at the S and Te atoms, 
\begin{equation}
    \sum_j q_j \ln{R_j} = q (\sum_{j\epsilon Te}\ln{R_j}-\sum_{j\epsilon S}\ln{R_j})
\end{equation}
As shown in \ref{fig:S2}, the potential at the center of the nanotube remains independent of z-values. If N is the number of S(or equivalently Te) atoms per unit cell of the nanotube, the $k=0$ part of the Eq.~(\ref{eq:V_final}) can be expressed as, 
\begin{equation}
V = -\frac{2K q N}{c}
\left(
\langle \ln R_j \rangle_{\mathrm{Te}}
-
\langle \ln R_j \rangle_{\mathrm{S}}
\right)
\label{eq:V_modif}
\end{equation}

Let R be the mean radius concerning each atom with the variance,
\begin{equation}
    \sigma^2 = \langle R_j^2\rangle -R^2
\end{equation}
Using the Taylor series expansion of \( \ln{R_j} \) around R,
\begin{equation}
    \langle \ln R_j \rangle = \ln{R} - \frac{\sigma^2}{2R^2} +...
\label{eq:lnR_approx} 
\end{equation}
For SW nanotubes, $\sigma \ll R \Rightarrow \sigma^2 / R^2 \ll 1$, thus first-order approximation gives an accurate description of the potential.So, using the first term of the Eq.~(\ref{eq:lnR_approx}), Eq.~(\ref{eq:V_modif}) can be modified as, 
\begin{equation}
    V = -\frac{2KqN}{c}\ln\frac{R_{Te}}{R_S}
\end{equation}
The radial quadrupole moment of the cylindrical system is defined as 
\begin{equation}
    Q = -\sum_j q_j R_j^2 = -qN(R_{Te}^2 - R_S^2)
    \label{eq:quadrupole_1}
\end{equation}
Using, Eq.~(\ref{eq:quadrupole_1}), 
\begin{equation}
    V = \frac{2KQ}{c(R_{Te}^2-R_S^2)}\ln{\frac{R_{Te}}{R_S}}
    \label{eq:V_Q_R}
\end{equation}
Each nanotube is assigned the radius $R_{Te}$ as R, diameter $D$ as $2R_{Te}$, and the difference $R_{Te}-R_s$ as thickness $d$, the Eq.~(\ref{eq:V_Q_R}) can be expressed as, 
\begin{equation}
    V = \frac{2KQ}{cd(D-d)}\ln{\frac{D}{D-2d}}
    \label{eq:V_Q_D_d}
\end{equation}
 Using 
\begin{equation}
    Q = pN(D-d)
\end{equation}
Where, $p=qd$, the dipole moment of the single pair $S-Te$,  Eq.~(\ref{eq:V_Q_D_d})  can be rewritten as
\begin{equation}
    V = \frac{2KpN}{cd}\ln{\frac{D}{D-2d}}
\end{equation}
For DW, Eq.~(\ref{eq:V_Q_D_d}) is evaluated separately for inner and outer tubes, and the potential is calculated by summing the contributions from both tubes.
\subsection{Calculation of effective charges}
The effective charges at S and Te atom locations are calculated by matching the $k=0$ part of the analytical formula, Eq.~(\ref{eq:V_final}) to the DFT calculated electrostatic potential, %at the center, 
\begin{equation}
    V_{\mathrm{DFT}} = -\frac{2K}{c} \sum_j q_j \ln{R_j}    
\end{equation}
The effective charge signs are assigned based on the relative electronegativity values, i.e. positive at Te and negative at S atoms. The magnitude of effective charges can be calculated as, 
\begin{equation}
q = \frac{V_{\mathrm{DFT}}\, c}
{2 K \sum_j q_j^{(\pm 1)} \ln R_j}
\label{eq:q_calcu}
\end{equation}
We use the same DCD model for 2D MoSTe, with positive and negative point charges located at the Te and S atoms, respectively to represent the out-of-plane dipoles. The effective charge ($q$) for the 2D MoSTe can be determined from,
\begin{equation}
    p = qd
\end{equation}
From DFT calculations, the dipole moment $p$ is 0.4838 Debye. The S-Te distance $d$ is 3.38 \AA{}. Therefore $q$ = 0.0298 e.\par
%\clearpage

%\bibliography{SI_ref}
\bibliography{nanotube}
%%%\clearpage